\documentclass[sort,numbers,journal=jacsat,manuscript=article]{achemso}
\SectionNumbersOn
\usepackage{tikz} 
\usetikzlibrary{quantikz2}

\usepackage{multicol}
\usepackage{graphicx}
\usepackage{graphicx}
\usepackage{dcolumn}
\usepackage{bm}
\usepackage{pgffor}
\usepackage[utf8]{inputenc}
\usepackage[T1]{fontenc}
\usepackage{mathptmx}
\usepackage{listings}
\usepackage{rotating} 
\usepackage{caption}
\usepackage{subcaption}

\usepackage{color}
\usepackage{dcolumn} 
\usepackage{bm} 
\usepackage{graphicx}
\usepackage{multirow} 
\usepackage{pifont} 
\usepackage{epsfig}
\usepackage{amsmath} 
\usepackage{float}
\usepackage{booktabs}
\usepackage{tabularx}
\usepackage{natbib}
\usepackage{gensymb}
\usepackage{rotating}
\usepackage{threeparttable}
\usepackage{comment}
\usepackage[normalem]{ulem}
\usepackage[hidelinks]{hyperref} 
\usepackage[version=3]{mhchem} 
\usepackage{xcolor}
\usepackage{amsmath,amssymb,amsthm}
\usepackage{mathtools,physics}
\usepackage{subcaption}

\usepackage{lineno}

\usepackage{xr-hyper}

\usepackage{xr}
\makeatletter

\newcommand*{\addFileDependency}[1]{
\typeout{(#1)}
\@addtofilelist{#1}
\IfFileExists{#1}{}{\typeout{No file #1.}}
}\makeatother

\newcommand*{\myexternaldocument}[1]{%
\externaldocument{#1}%
\addFileDependency{#1.tex}%
\addFileDependency{#1.aux}%
}

\myexternaldocument{SI}
\newcommand{\siref}[1]{S\ref{#1}}

\title{Parametrized Quantum Circuit Learning for Quantum Chemical Applications}
\author{Grier M. Jones}
\affiliation[UTSG ECE]{
The Edward S. Rogers Sr. Department of Electrical and Computer Engineering, 
University of Toronto, 
10 King's College Road, Toronto, Ontario, 
Canada M5S 3G4}
\alsoaffiliation[UTM CHEM]{
Department of Chemical and Physical Sciences, 
University of Toronto Mississauga, 
3359 Mississauga Road, Mississauga, Ontario, 
Canada L5L 1C6}
\alsoaffiliation[UTM CHEM]{
Department of Chemical and Physical Sciences, 
University of Toronto Mississauga, 
3359 Mississauga Road, Mississauga, Ontario, 
Canada L5L 1C6}
\alsoaffiliation[GMJ ORCID]{https://orcid.org/0000-0003-0982-3129}
\email{grier.jones@utoronto.edu}
          
\author{Viki Kumar Prasad}
\affiliation[UTSG ECE]{
The Edward S. Rogers Sr. Department of Electrical and Computer Engineering, 
University of Toronto, 
10 King's College Road, Toronto, Ontario, 
Canada M5S 3G4}
\alsoaffiliation[UTM CHEM]{
Department of Chemical and Physical Sciences, 
University of Toronto Mississauga, 
3359 Mississauga Road, Mississauga, Ontario, 
Canada L5L 1C6}
\alsoaffiliation[UT DSI]{
	Data Sciences Institute, 
	University of Toronto, 
	700 University Ave 10th floor,
	 Toronto, ON M7A 2S4}
\alsoaffiliation[UCalg]{Current Affiliation: Department of Chemistry, 
	University of Calgary, 
	2500 University Drive NW,
	 Calgary, AB T2N 1N4}
\alsoaffiliation[VKP ORCID]{https://orcid.org/0000-0003-0982-3129}

\author{Ulrich Fekl}
\affiliation[UTM CHEM]{
Department of Chemical and Physical Sciences, 
University of Toronto Mississauga, 
3359 Mississauga Road, Mississauga, Ontario, 
Canada L5L 1C6}
\alsoaffiliation[UF ORCID]{https://orcid.org/0000-0001-7345-1716}

\author{Hans-Arno~Jacobsen}
\affiliation[UTSG ECE]{
The Edward S. Rogers Sr. Department of Electrical and Computer Engineering, 
University of Toronto, 
10 King's College Road, Toronto, Ontario, 
Canada M5S 3G4}
\alsoaffiliation[UTSG CS]{
Department of Computer Science, 
University of Toronto, 
40 St. George Street, Toronto, Ontario, 
Canada M5S 2E4}
\alsoaffiliation[HAJ ORCID]{https://orcid.org/0000-0003-0813-0101}
\abbreviations{}

\begin{document}

\section*{Abstract}
In the field of quantum machine learning (QML), parametrized quantum circuits (PQCs)—constructed using a combination of fixed and tunable quantum gates—provide a promising hybrid framework for tackling complex machine learning problems. Despite numerous proposed applications, there remains limited exploration of datasets relevant to quantum chemistry.
In this study, we investigate the potential benefits and limitations of PQCs on two chemically meaningful datasets: (1) the BSE49 dataset, containing bond separation energies for 49 different classes of chemical bonds, and (2) a dataset of water conformations, where coupled-cluster singles and doubles (CCSD) wavefunctions are predicted from lower-level electronic structure methods using the data-driven coupled-cluster (DDCC) approach.
We construct a comprehensive set of 168 PQCs by combining 14 data encoding strategies with 12 variational ans{\"a}tze, and evaluate their performance on circuits with 5 and 16 qubits. Our initial analysis examines the impact of circuit structure on model performance using state-vector simulations. We then explore how circuit depth and training set size influence model performance.
Finally, we assess the performance of the best-performing PQCs on current quantum hardware, using both noisy simulations (``fake'' backends) and real quantum devices. Our findings underscore the challenges of applying PQCs to chemically relevant problems that are straightforward for classical machine learning methods but remain non-trivial for quantum approaches.\par

\newpage
\section{Introduction}
In recent years, machine learning (ML) has emerged as a popular tool in chemistry to reveal new patterns in data, provide new insights beyond simple models, accelerate computations, and analyze chemical space.
For computational chemists, the primary goal of applying ML is often to circumvent the explicit calculation of molecular properties, which can be computationally expensive for large datasets.\cite{janet_machine_2020}
ML can be applied to a diverse set of problems including, but not limited to, accelerating molecular simulations\cite{behler_perspective_2016,ssmith_ani-1_2017,gao_torchani_2020}, determining molecular properties\cite{yang_analyzing_2019,ramakrishnan_quantum_2014,ramakrishnan_big_2015,hansen_machine_2015,unke_physnet_2019}, and for discovering new catalysts\cite{zhong_accelerated_2020,nandy_computational_2021,mjones_data-driven_2023}, drugs\cite{goh_deep_2017,yang_concepts_2019}, and materials.\cite{butler_machine_2018,sanchez-lengeling_inverse_2018,raccuglia_machine-learning-assisted_2016}
Since these applications can become resource intensive, regarding the generation of training data using traditional computational chemistry approaches and the training of large-scale ML models, computational chemistry and ML practitioners have explored new techniques and acceleration platforms, such as graphical processing units (GPUs) and tensor processing units (TPUs).\cite{prasad2024bridging,ufimtsev_graphical_2008,gotz_chapter_2010,pederson_large_2023,goh_deep_2017,gawehn_advancing_2018,pandey_transformational_2022,ssmith_ani-1_2017}

Alternatively, computational approaches that leverage the quantum mechanical principles of superposition and entanglement—collectively known as quantum computing (QC)—are gaining popularity in chemical applications, driven by the potential for quantum speedups in quantum chemical calculations.\cite{cao_quantum_2019}
In computational chemistry, phase estimation-based quantum algorithms have been theoretically proven to yield exponential speedups over classical exact methods such as full configuration interaction.\cite{abrams_simulation_1997,abrams_quantum_1999,aspuru-guzik_simulated_2005,lanyon_towards_2010,whitfield_simulation_2011,aspuru-guzik_photonic_2012}
Despite the promising speedups, QPE requires long coherence times, while the current generation of quantum processing units (QPUs) is often too noisy for practical applications.
Alternatively, methods based on the variational principle, such as the variational quantum eigensolver (VQE)\cite{peruzzo_variational_2014,cerezo_variational_2021,mcclean_theory_2016,bharti_noisy_2022}, have been proposed as a quantum--classical hybrid approach, capable of running on noisy, near-term quantum devices.

While most QC studies that are relevant to computational chemists focus on creating more efficient electronic structure methods on quantum computers\cite{romero_strategies_2019,mcardle_quantum_2020,bauer_quantum_2020,cao_quantum_2019}, an approach that combines both ML and QC is quantum machine learning (QML).
Using either formal mathematical proofs or numerical results based on empirical observations, QML has shown potential quantum speedups for various applications using a diverse set of implementations.\cite{biamonte_quantum_2017}
While several classes of QML algorithms have shown promise for providing flexible ML models, parametrized quantum circuits (PQCs) can achieve non-trivial results on near-term quantum hardware.
PQCs formulate the ML algorithm as a variational problem, optimized using a hybrid approach that combines classical and quantum hardware.\cite{benedetti_parameterized_2019}
Like classical ML approaches, PQCs have been applied to several chemistry use cases such as drug\cite{suzuki_predicting_2020,smaldone_quantum--classical_2024,bhatia_quantum_2023,kao_exploring_2023,li_quantum_2021,avramouli_quantum_2023,avramouli_unlocking_2023} and materials discovery\cite{ishiyama_noise-robust_2022,ryu_quantum_2023,vitz_hybrid_2024}, the prediction of proton affinities\cite{jin_integrating_2025}, and experimental molecular properties, including the log solubility in water, melting point, octanol/water distribution coefficient, and hydration free energy of small molecules in water.\cite{hatakeyama-sato_quantum_2023}
Despite the broad range of topics covered in these studies and the interest among computational chemists in exploring PQCs for chemical applications, there is a lack of studies analyzing the potential benefits or drawbacks of using QML for quantum chemistry. 

In this study, we analyze a diverse set of PQCs using two datasets generated using quantum chemical methods.
The first dataset, BSE49, consists of bond separation energies (BSEs) of 49 unique bond types, calculated using the highly accurate (RO)CBS-QB3 composite method.\cite{prasad_bse49_2021}
The second dataset consists of water conformers calculated with coupled-cluster singles and doubles (CCSD) using the data-driven coupled-cluster (DDCC) scheme of Townsend and Vogiatzis.\cite{townsend_data-driven_2019,jones_chapter_2023}
Both datasets offer a unique perspective on the aptitude of applying PQCs on classical and quantum data\cite{cerezo_challenges_2022} since the models based on BSE49 rely on classical molecular representations\cite{jones_molecular_2023}, such as Molecular ACCess Systems (MACCS)\cite{durant_reoptimization_2002} or Morgan fingerprints \cite{morgan_generation_1965,rogers_extended-connectivity_2010}, as input, while the input features in the DDCC method encode explicit quantum information related to the molecular electronic structure.
To help facilitate the exploration of PQCs for regression-based QML tasks, we introduce \textsc{qregress}, a modular Python framework based on \textsc{\textsc{PennyLane}}\cite{bergholm_pennylane_2022} and \textsc{\textsc{Qiskit}}\cite{javadi-abhari_quantum_2024}.
To this end, we investigate the effects of classical and quantum data using a comprehensive set of 168 unique PQCs, which are constructed from a combination of 14 data encoding and 12 variational layers.
We then perform an analysis of circuit depth on model performance, using two different expansion strategies, one based on data re-uploading\cite{perez-salinas_data_2020} and the other based on increasing the number of model parameters using additional variational layers.
Using these insights, we then analyze how the best model performs using noisy simulators and real quantum hardware.
Lastly, we provide a detailed discussion on the efficiency and performance of PQCs, with insights into what quantum enhancement could mean in comparison to classical ML models.

\section{Methods}
\subsection{Parametrized Quantum Circuits}
PQCs typically consist of three components: encoding layers that map features onto a quantum circuit, variational layers with classically optimized parameters, and measurement operations that provide numerical estimates of the regression target values.\cite{suzuki_predicting_2020} 
Choosing the optimal encoding layer can be a challenging task due to the costs associated with mapping the input data to qubits.\cite{biamonte_quantum_2017} 
Due to this fact, we choose existing encoding layers that have shown promising results for regression tasks such as Mitarai (M)\cite{mitarai_quantum_2018}, single-angle (A1) and double-angle (A2) encoding layers\cite{suzuki_predicting_2020}, along with the instantaneous quantum polynomial (IQP) circuit.\cite{bremner_average-case_2016,havlicek_supervised_2019}

Note that in the following section, we adopt the notation introduced in the paper of \citet{suzuki_predicting_2020} to ensure consistency.
Encoding layers work by mapping a given $d$-dimensional feature vector, $\mathbf{x}=(x_{1}, x_{2}, \ldots, x_{d})^{T} \in \mathbb{R}^{d}$, normalized on the range $[-1,1]$, onto a quantum circuit using a unitary matrix, denoted as $U_{\Phi(\mathbf{x})}$, to produce the quantum state $U_{\Phi(\mathbf{x})}\ket{0}^{\otimes n}$, where $n$ is the number of qubits.
More generally, the encoding layer can also incorporate both single-qubit and entanglement gates, such that 
\begin{equation}
	U_{\Phi(\mathbf{x})} =  \prod_{l} E_{\text{ent}}^{l} U_{\phi_{l}(\mathbf{x})},
	\label{eq:general_encoding}
\end{equation}
where, $E_{\text{ent}}^{l}$ denotes the entangling gates, e.g., CNOT or CZ, and  $U_{\phi_{l}(\mathbf{x})}$ denotes the choice of encoding unitaries. 

\begin{figure}[H]
	\centering
	\begin{subfigure}[b]{0.3\textwidth}
		\centering
		\includegraphics[width=0.5\textwidth]{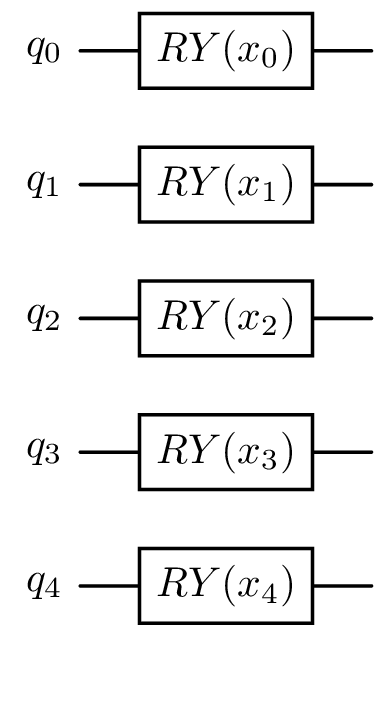}
		\caption{}
		\label{fig:A1}
	\end{subfigure}
	\hfill
	\begin{subfigure}[b]{0.3\textwidth}
		\centering
		\includegraphics[width=0.8\textwidth]{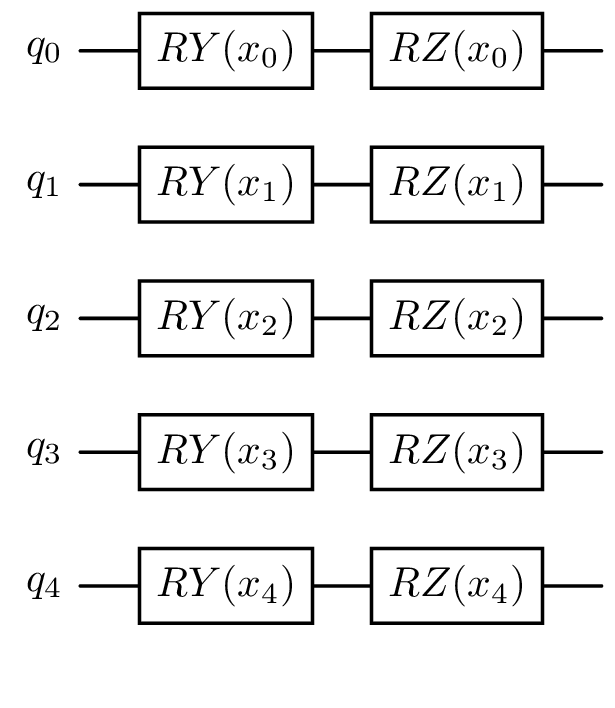}
		\caption{}
		\label{fig:A2}
	\end{subfigure}
	\hfill
	\begin{subfigure}[b]{0.3\textwidth}
		\centering
		\includegraphics[width=1.1\textwidth]{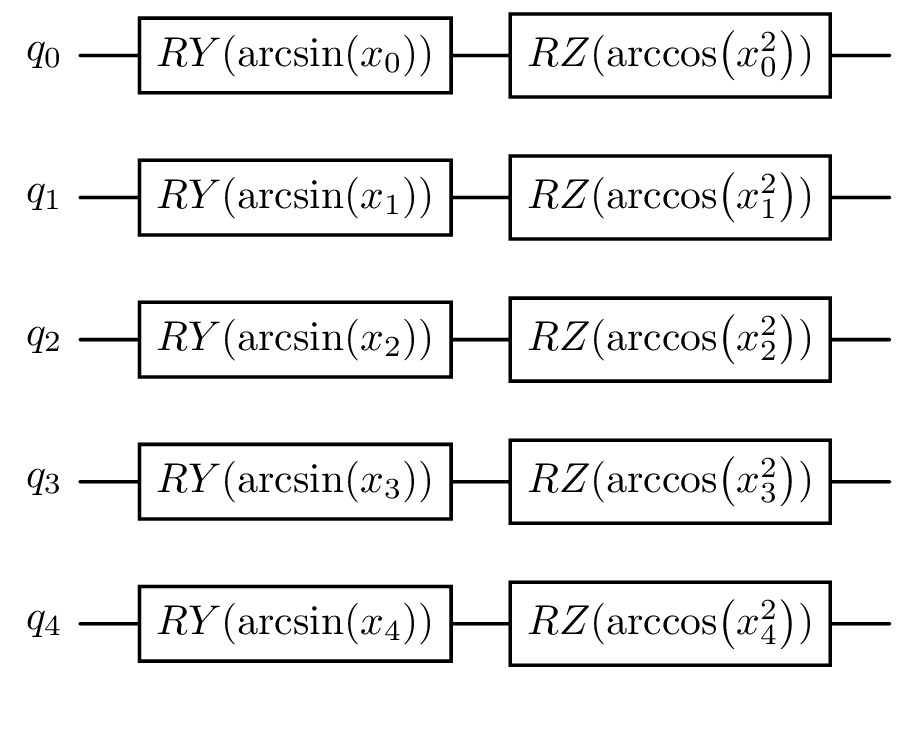}
		\caption{}
		\label{fig:M}
	\end{subfigure}
	\hfill
	\begin{subfigure}[b]{\textwidth}
		\centering
		\includegraphics[width=\textwidth]{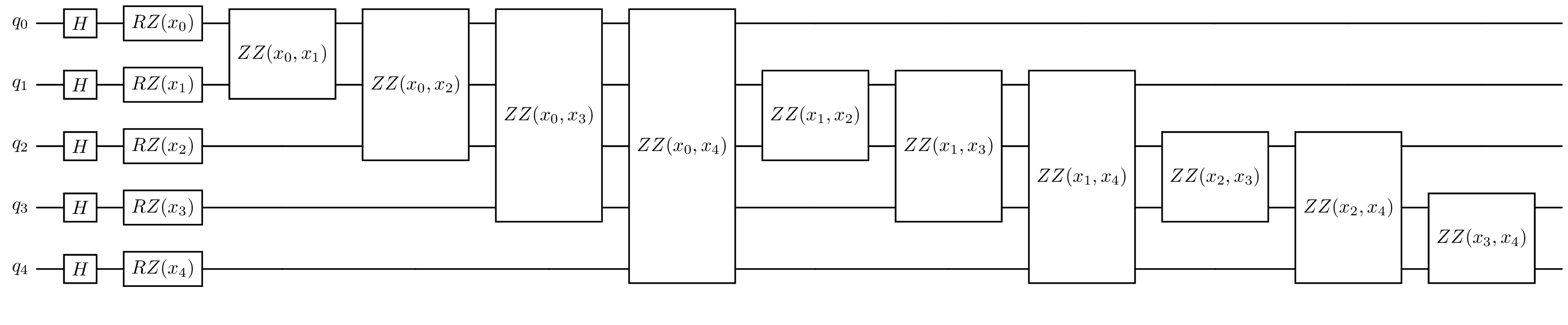}
		\caption{}
		\label{fig:IQP}
	\end{subfigure}
	\hfill    
	\caption{Examples of the encoding layers, using five qubits, analyzed in this study: (a) single-angle (A1), (b) double-angle (A2), (c) Mitarai (M), and (d) Instantaneous Quantum Polynomial (IQP).}
	\label{fig:encoders}
\end{figure}

In this study, like the study of \citet{suzuki_predicting_2020}, we explore encoders with $l={1,2}$. 
When $l=1$, this forms the simplest encoding layers with $E_{\text{ent}}^{1}$ being absent, e.g., corresponding to an identity matrix, $\mathbf{I}$. 
Using the previously defined notation, the simplest encoding layer, or single-angle encoding (Fig.~\ref{fig:encoders} \textbf{(a)}), is defined as,
\begin{equation}
	U_{\text{A1}} = \prod_{i=0}^{n-1} RY_{i}(x_{i}),
	\label{eq:A1}
\end{equation}
where $RY_{i}$ denotes a parametrized $RY$ rotation gate on qubit $i$.
Single-angle encoding, or qubit encoding, is structured similarly to a product of unentangled quantum states and has a similar mathematical structure as a product of local kernels where each $x_{i}$ is encoded in a local feature map.\cite{stoudenmire_supervised_2016,benedetti_parameterized_2019}

The next encoder, double-angle encoding (Fig.~\ref{fig:encoders} \textbf{(b)}) utilizes a parametrized $RY$ rotation gate on qubit $i$, similar to $A1$, with the addition of a parametrized $RZ$ rotation gate on qubit $i$, denoted as
\begin{equation}
	U_{\text{A2}} = \prod_{i=0}^{n-1}  RZ_{i}(x_{i}) RY_{i}(x_{i}).
	\label{eq:A2}
\end{equation}
The double-angle encoding introduces additional redundancy by encoding two angles on the Bloch sphere.
An extension of the double-angle encoding is the Mitarai encoding layer (Fig.~\ref{fig:encoders} \textbf{(c)}), which includes an arccosine function on the parametrized $RZ$ gate and arcsine on the parametrized $RY$ gate,
\begin{equation}
	U_{\text{M}}  = \prod_{i=0}^{n-1} RZ_{i}(\arccos (x_{i}^{2})) RY_{i}(\arcsin (x_{i})).
	\label{eq:M}
\end{equation}
This unitary is physically motivated by expanding a density operator in terms of a set of Pauli operators.\cite{mitarai_quantum_2018}

The last and most complex encoding layer is the instantaneous quantum polynomial (IQP) (Fig.~\ref{fig:encoders}~\textbf{(d)}) proposed by Havlicek \textit{et al.}\cite{havlicek_supervised_2019},
\begin{equation}
	U_{\text{IQP}}  = \prod_{i<j} ZZ_{ij}\prod_{i=0}^{n-1} RZ_{i}(x_{i})H_{i},
	\label{eq:IQP}
\end{equation}
where $H_{i}$ denotes a Hadamard gate on qubit $i$ and $ZZ_{ij}$ denotes a two-qubit entangling gate defined as $ZZ_{ij} = e^{-i x_{i} x_{j} \sigma_{z} \otimes \sigma_{z}}$.
It should be noted that under specific complexity and theoretic assumptions, IQP circuits cannot be efficiently simulated using classical resources and, therefore, offer a circuit that can only be simulated efficiently using quantum resources.\cite{lund_quantum_2017,harrow_quantum_2017}

When $l=2$, more complex circuits can be generated when we choose the entanglement gates, $E_{\text{ent}}^{1}$ and $E_{\text{ent}}^{2}$, to be equivalent and the encoding layer takes the following form, $U_{\Phi(x)} =  E_{\text{ent}} U_{\phi_{2}(\mathbf{x})} E_{\text{ent}} U_{\phi_{1}(\mathbf{x})}$.
Additionally, we exclude IQP encoding when $l=2$ due to the increased computational costs associated with the circuit depth when compared to the expanded models based on A1, A2, and M encoding.
Therefore, when we exclude IQP, there are 5 unique combinations of $U_{\phi_{1}(\mathbf{x})}$ and $U_{\phi_{2}(\mathbf{x})}$ (M-M, A1-A1, A2-A2, M-A1, and M-A2) and two different entanglement layer options (CNOT and CZ) for a total of 10 encoding circuits with $l=2$.
The common notation for these circuits is $U_{\phi_{1}(\mathbf{x})}-U_{\phi_{2}(\mathbf{x})}-E_{\text{ent}}$, where two example encoding circuits are M--M--CNOT and M--A1--CNOT.
The 14 encoding circuits used throughout this study are shown in Table \ref{tab:encoders}, where the first column shows the label, second column the first unitary ($U_{\phi_{1}(\mathbf{x})}$), third column the second unitary ($U_{\phi_{2}(\mathbf{x})}$), and last column the entanglement gates ($E_{\text{ent}}$).

\begin{table}[htbp]
	\centering
	\begin{tabular}{|c|c|c|c|}
		\hline
		\textbf{Name} & $U_{\phi_{1}(\mathbf{x})}$ & $U_{\phi_{2}(\mathbf{x})}$ & $E_{\text{ent}}$  \\
		\hline
		\hline
		A1 & $U_{\text{A1}}$ & --- & --- \\
		\hline
		A2 & $U_{\text{A2}}$ & --- & --- \\
		\hline		
		M & $U_{\text{M}}$ & --- & --- \\
		\hline
		IQP & $U_{\text{IQP}}$ & --- & --- \\
		\hline
		A1--A1--CNOT & $U_{\text{A1}}$ & $U_{\text{A1}}$ & $E_{\text{CNOT}}$ \\
		\hline
		 A2--A2--CNOT & $U_{\text{A2}}$ & $U_{\text{A2}}$ & $E_{\text{CNOT}}$ \\
		\hline
		M--M--CNOT & $U_{\text{M}}$ & $U_{\text{M}}$ & $E_{\text{CNOT}}$ \\
		\hline
		M--A1--CNOT & $U_{\text{M}}$ & $U_{\text{A1}}$ & $E_{\text{CNOT}}$ \\
		\hline		
		M--A2--CNOT & $U_{\text{M}}$ & $U_{\text{A2}}$ & $E_{\text{CNOT}}$ \\
		\hline				
		A1--A1--CZ & $U_{\text{A1}}$ & $U_{\text{A1}}$ & $E_{\text{CZ}}$ \\
		\hline
		A2--A2--CZ& $U_{\text{A2}}$ & $U_{\text{A2}}$ & $E_{\text{CZ}}$ \\
		\hline
		M--M--CZ & $U_{\text{M}}$ & $U_{\text{M}}$ & $E_{\text{CZ}}$ \\
		\hline
		M--A1--CZ & $U_{\text{M}}$ & $U_{\text{A1}}$ & $E_{\text{CZ}}$ \\
		\hline		
		M--A2--CZ & $U_{\text{M}}$ & $U_{\text{A2}}$ & $E_{\text{CZ}}$ \\
		\hline						
	\end{tabular}
	\caption{The 14 encoding layers used throughout this study. The first column shows the encoding circuit name, the second column the unitary for $l=1$, the third column the unitary for $l=2$, and the last column the entanglement gates.}
	\label{tab:encoders}
\end{table}

\begin{figure}[H]
	\centering
	\begin{subfigure}[hc]{0.49\textwidth}
		\centering
		\includegraphics[width=0.6\textwidth]{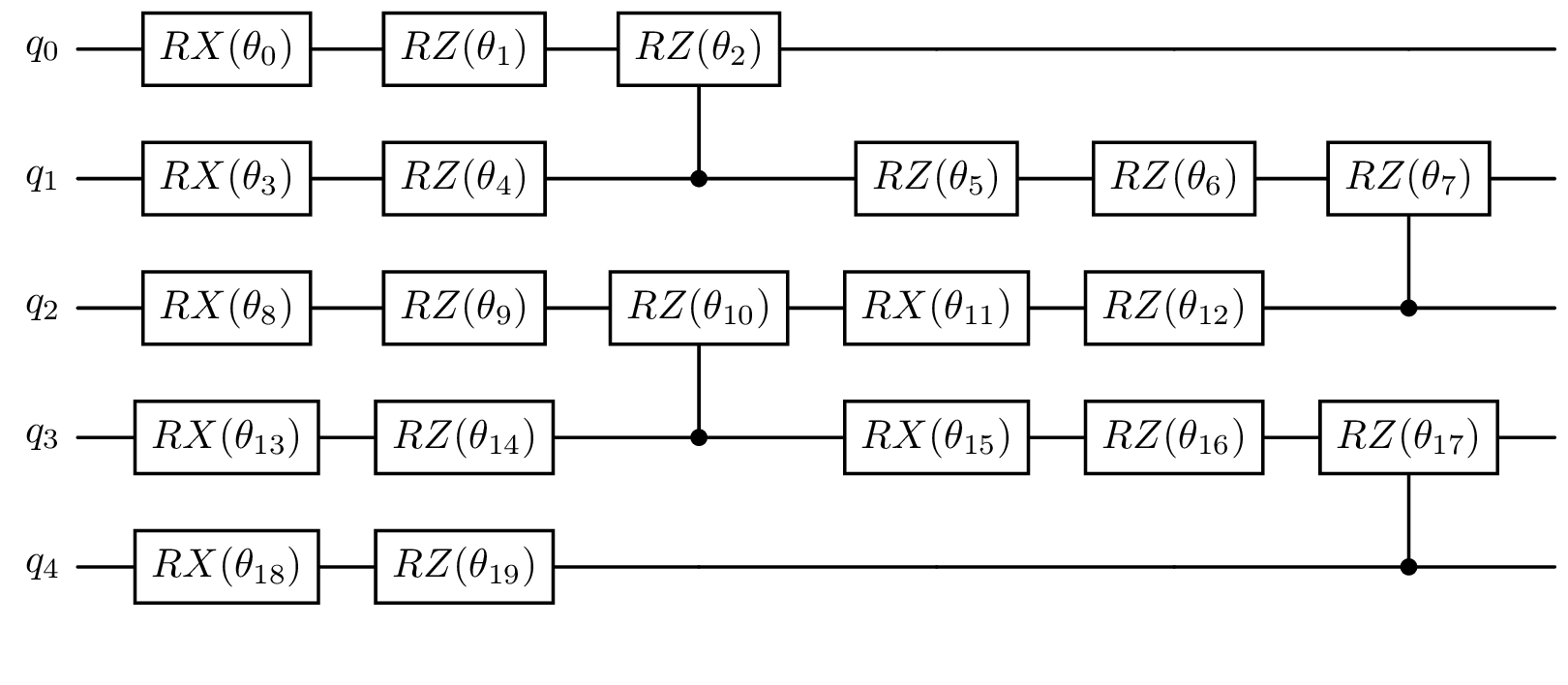}
		\caption{}
		\label{fig:Modified-Pauli-CRZ}
	\end{subfigure}
	\hfill
	\begin{subfigure}[hc]{0.49\textwidth}
		\centering
		\includegraphics[width=0.6\textwidth]{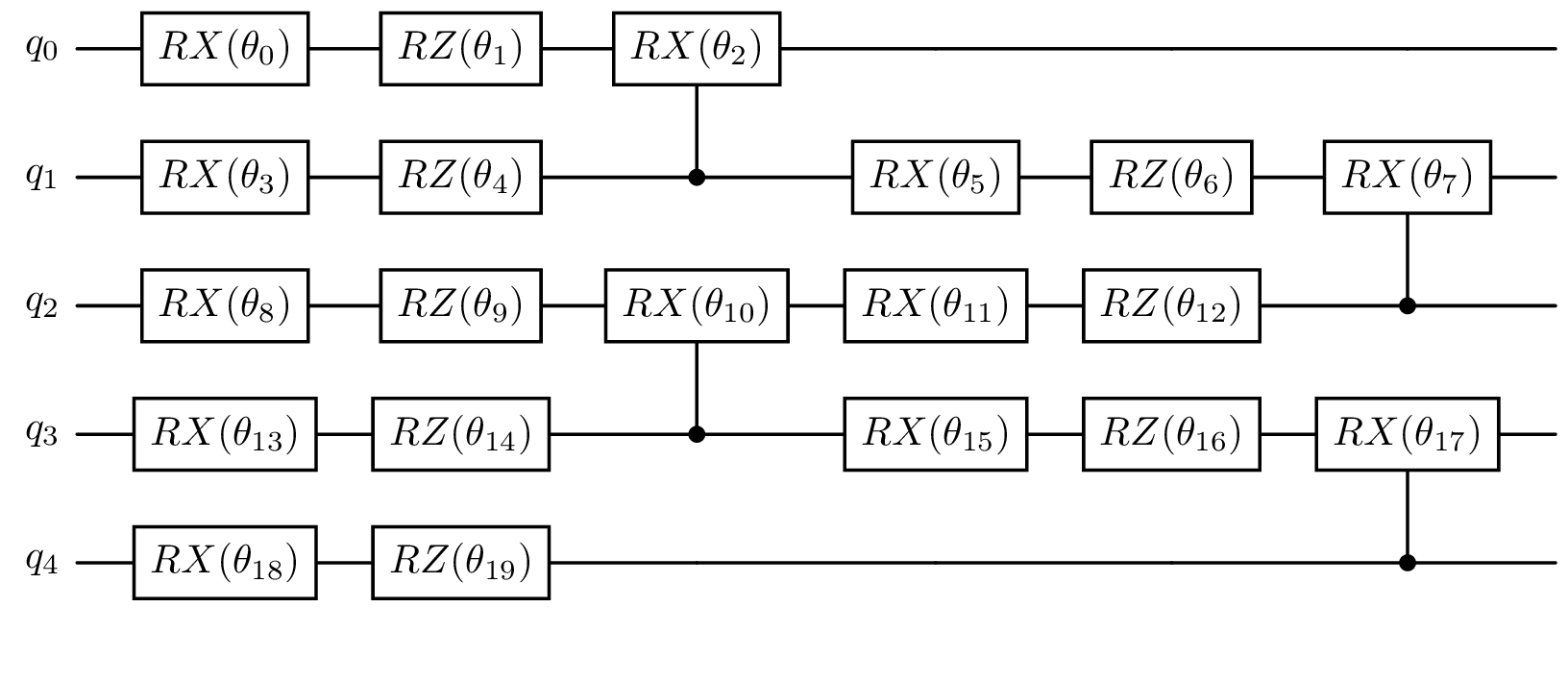}
		\caption{}
		\label{fig:Modified-Pauli-CRX}
	\end{subfigure}
	\hfill
	\begin{subfigure}[hc]{0.49\textwidth}
		\centering
		\includegraphics[width=0.6\textwidth]{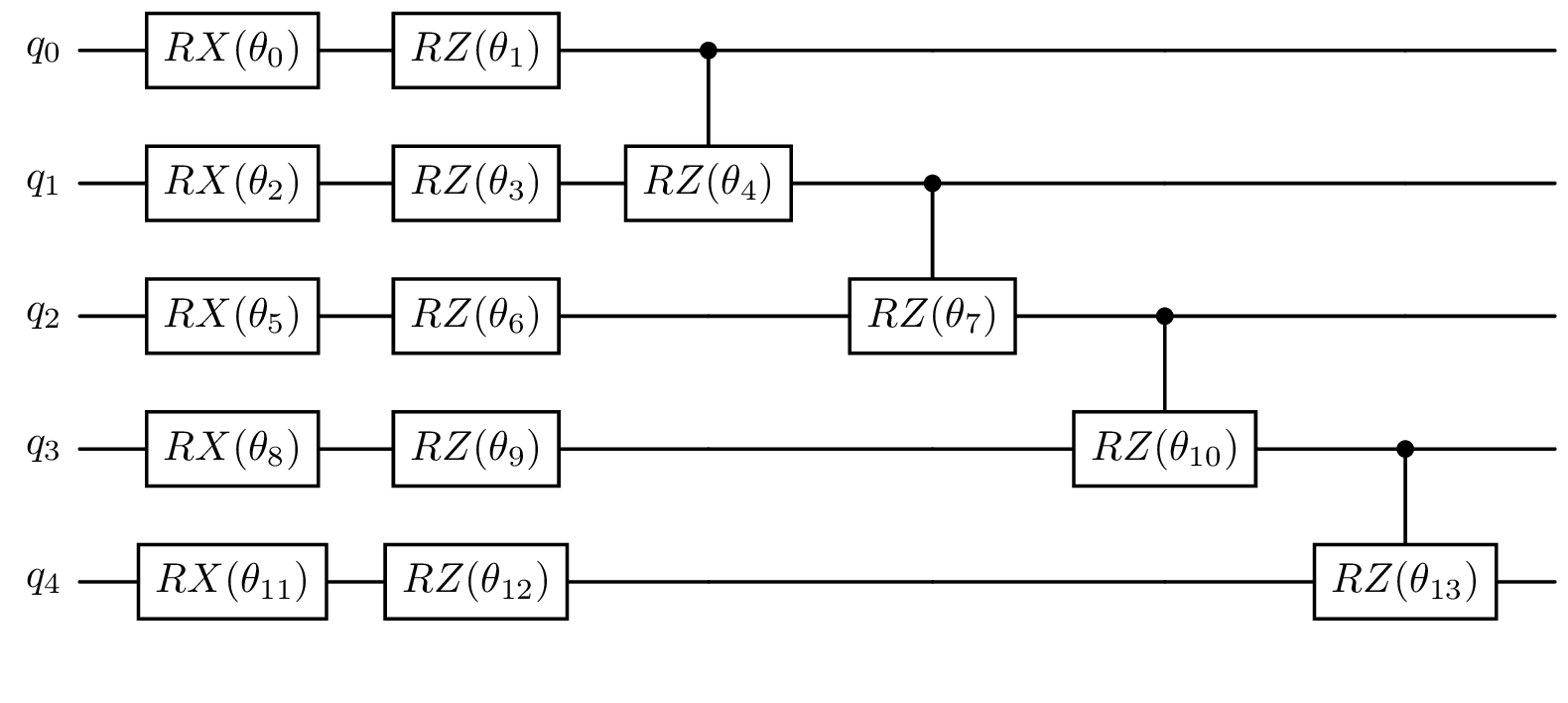}
		\caption{}
		\label{fig:Efficient-CRZ}
	\end{subfigure}
	\hfill
	\begin{subfigure}[hc]{0.49\textwidth}
		\centering
		\includegraphics[width=0.6\textwidth]{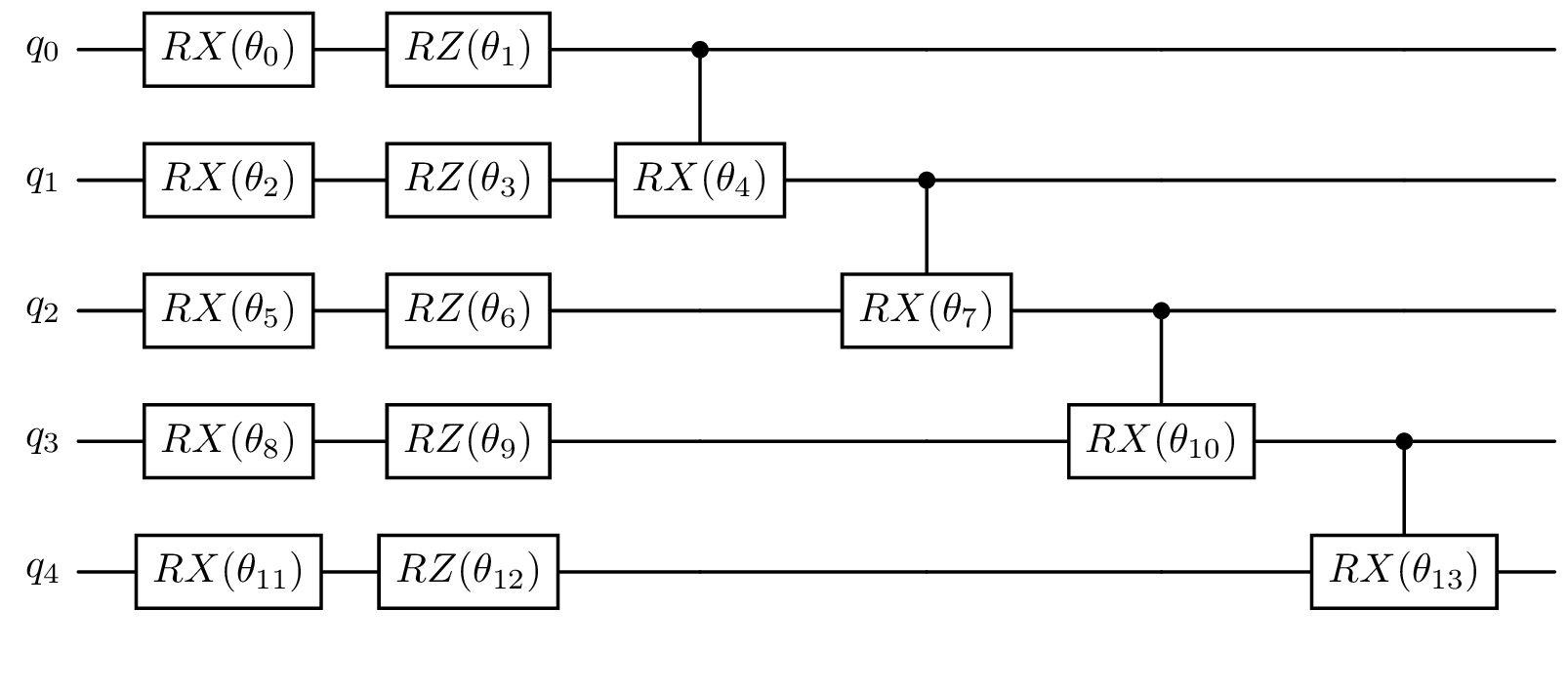}
		\caption{}
		\label{fig:Efficient-CRX}
	\end{subfigure}
	\hfill    
	\begin{subfigure}[hc]{0.49\textwidth}
		\centering
		\includegraphics[width=0.6\textwidth]{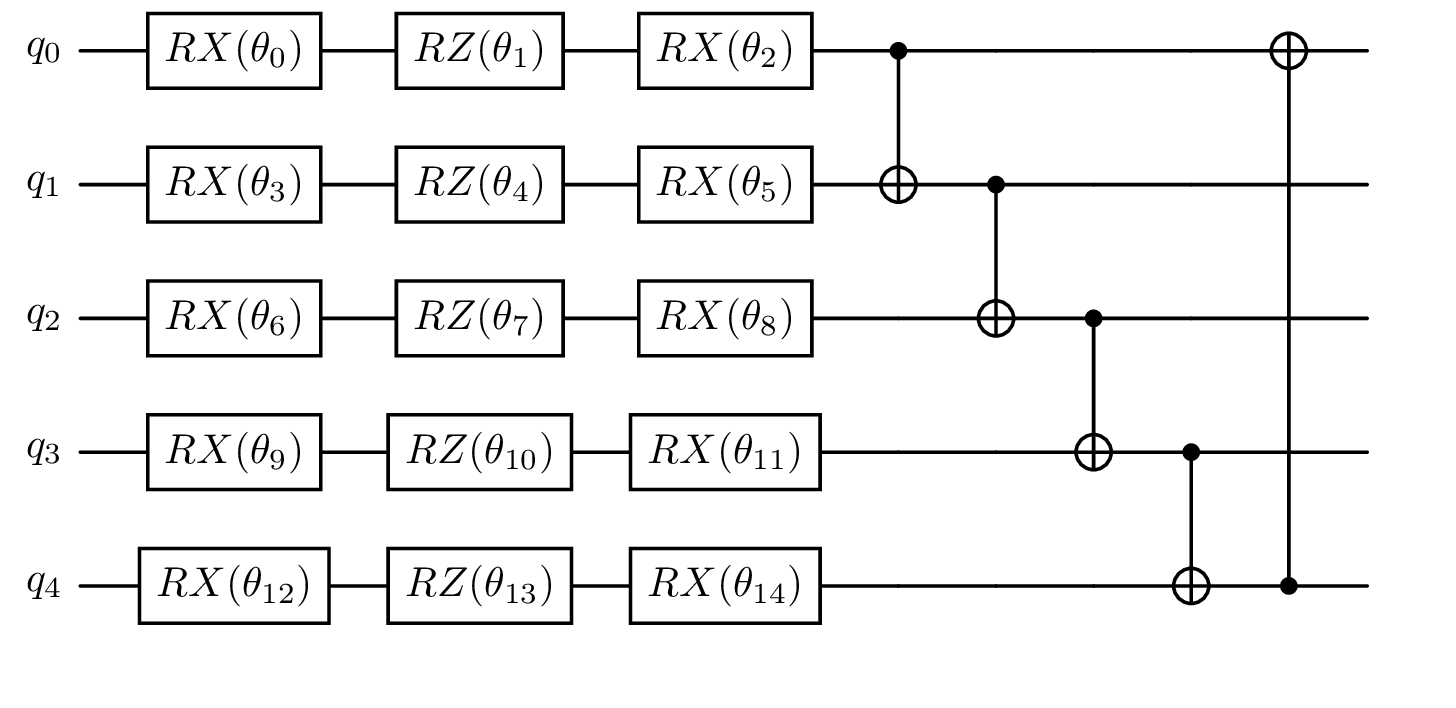}
		\caption{}
		\label{fig:HWE-CNOT}
	\end{subfigure}
	\hfill
	\begin{subfigure}[hc]{0.49\textwidth}
		\centering
		\includegraphics[width=0.6\textwidth]{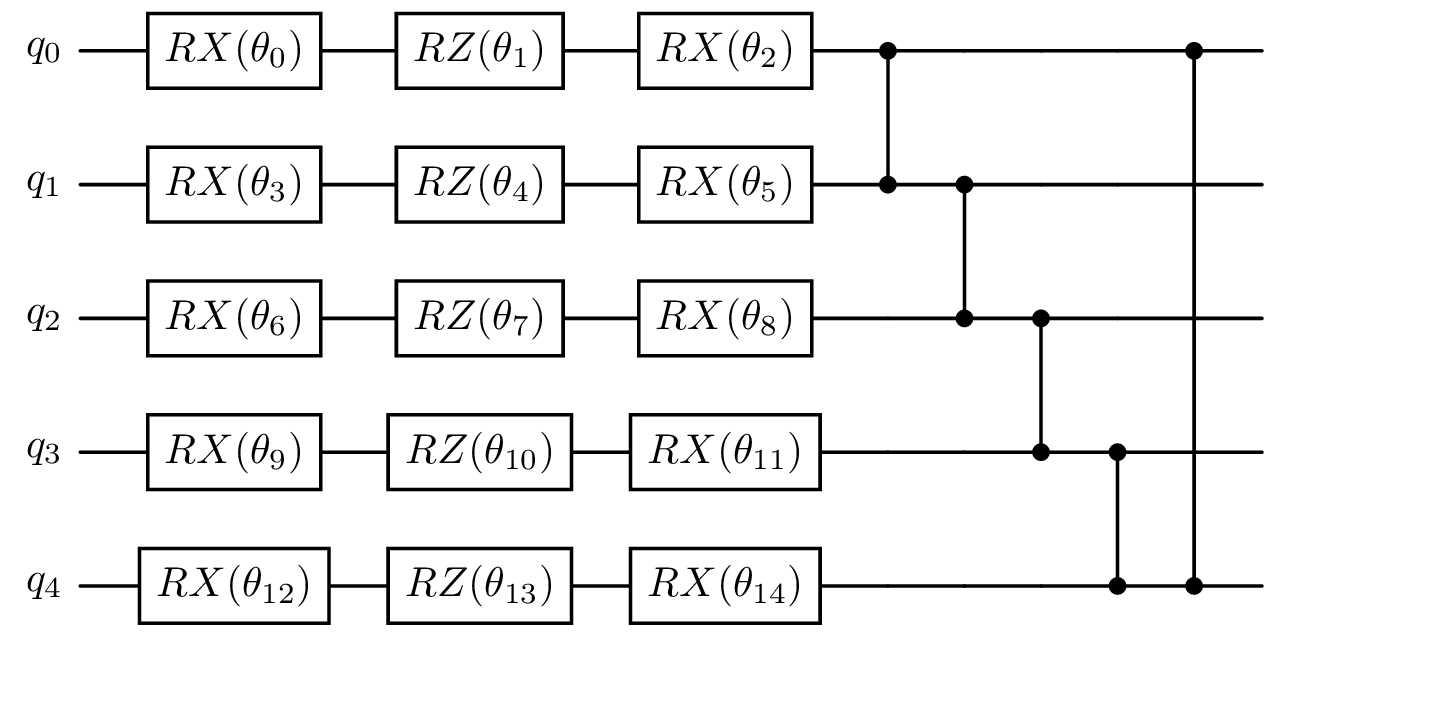}
		\caption{}
		\label{fig:HWE-CZ}
	\end{subfigure}
	\hfill
	\begin{subfigure}[hc]{0.49\textwidth}
		\centering
		\includegraphics[width=0.6\textwidth]{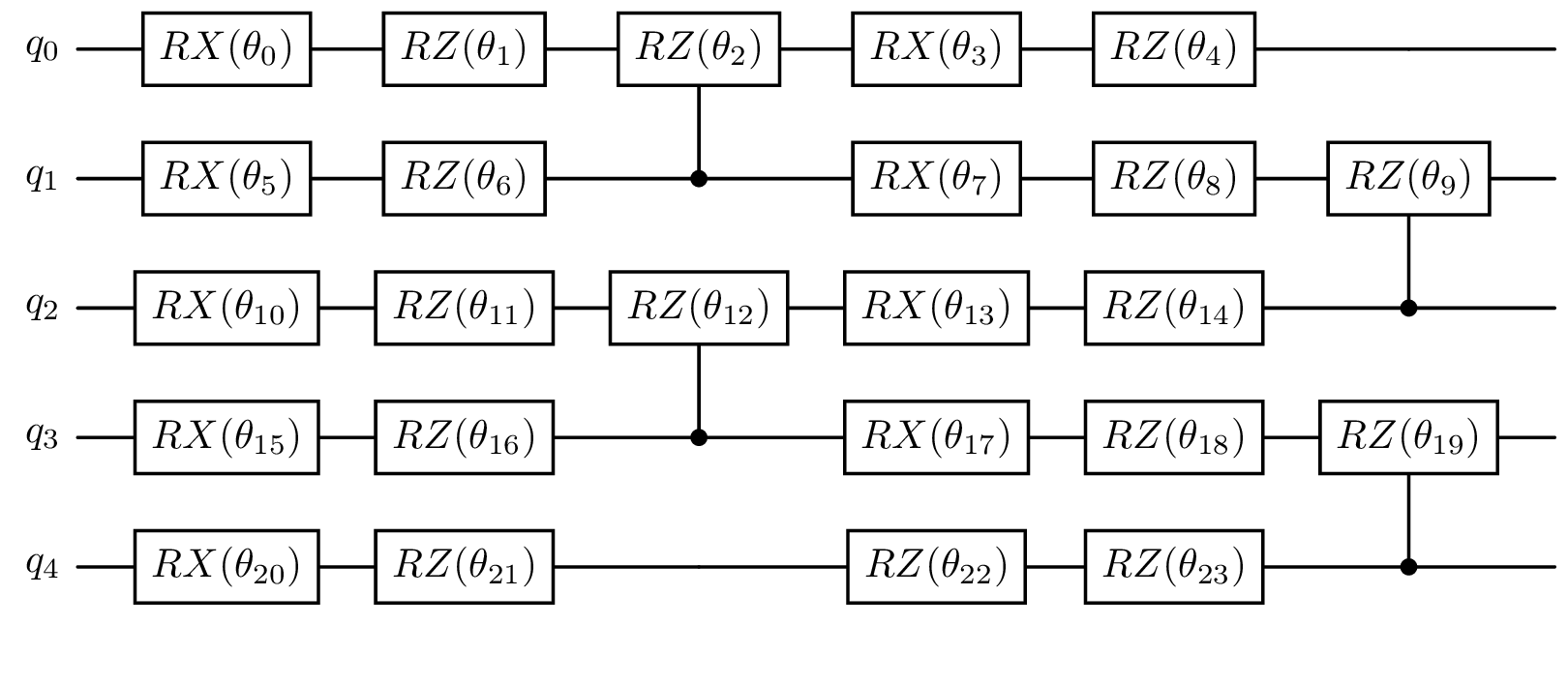}
		\caption{}
		\label{fig:Full-Pauli-CRZ}
	\end{subfigure}
	\hfill  
	\begin{subfigure}[hc]{0.49\textwidth}
		\centering
		\includegraphics[width=0.6\textwidth]{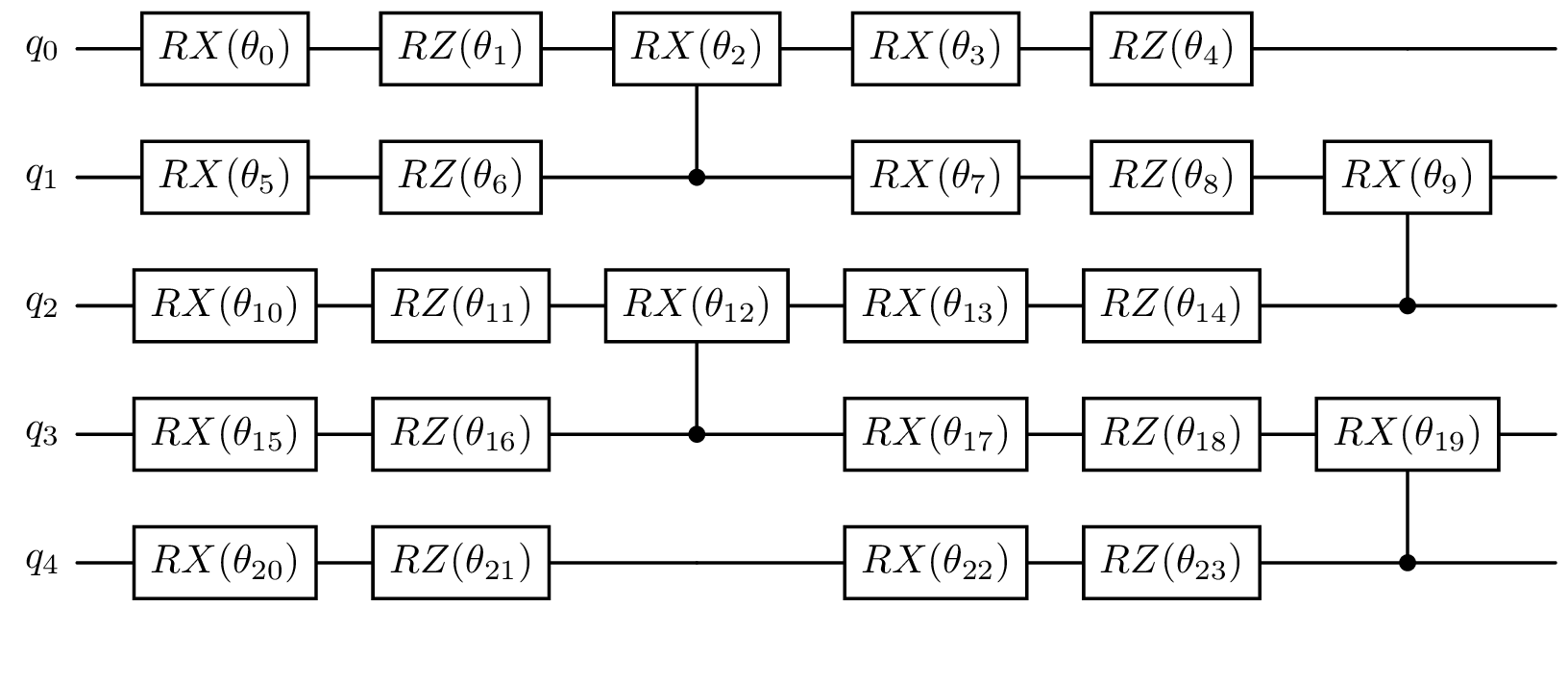}
		\caption{}
		\label{fig:Full-Pauli-CRX}
	\end{subfigure}
	\hfill
        \begin{subfigure}[hc]{0.49\textwidth}
		\centering
		\includegraphics[width=0.6\textwidth]{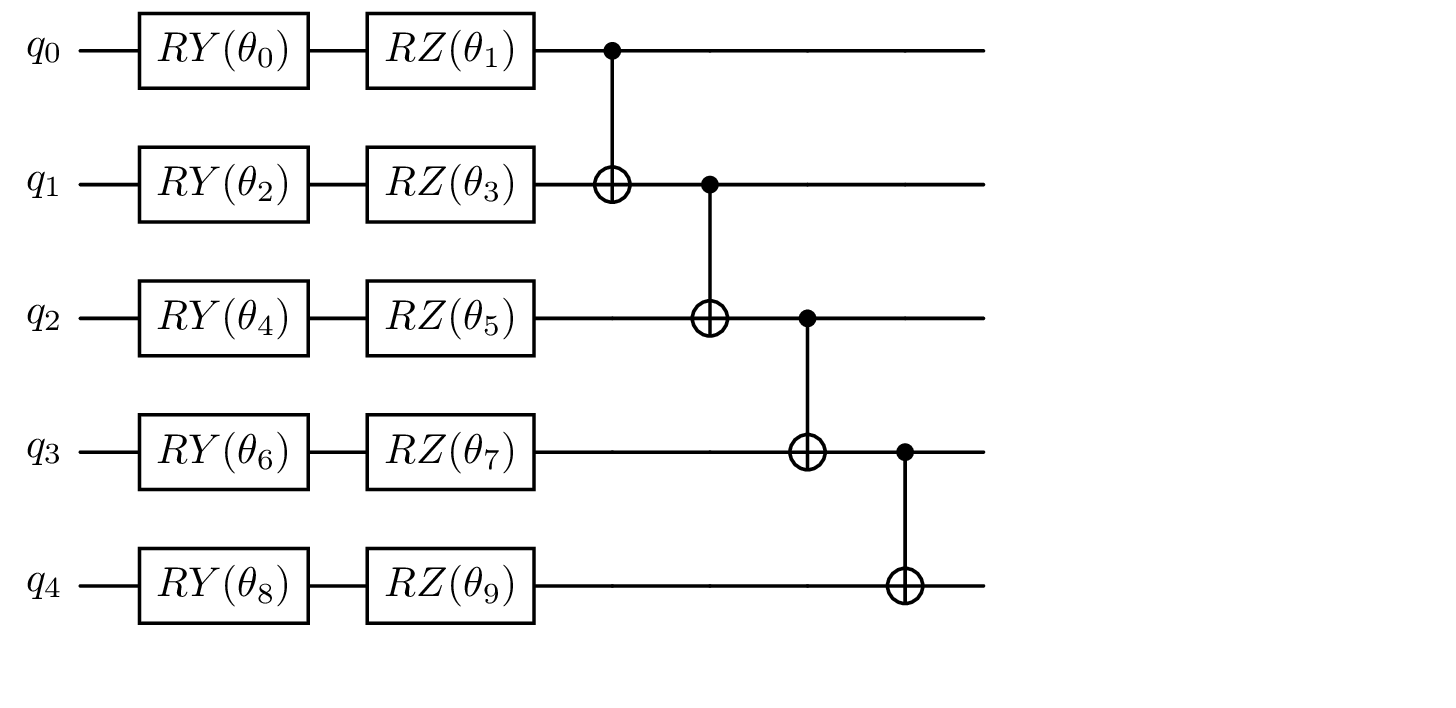}
		\caption{}
		\label{fig:ESU2}
	\end{subfigure}
	\hfill
	\begin{subfigure}[hc]{0.49\textwidth}
		\centering
		\includegraphics[width=0.6\textwidth]{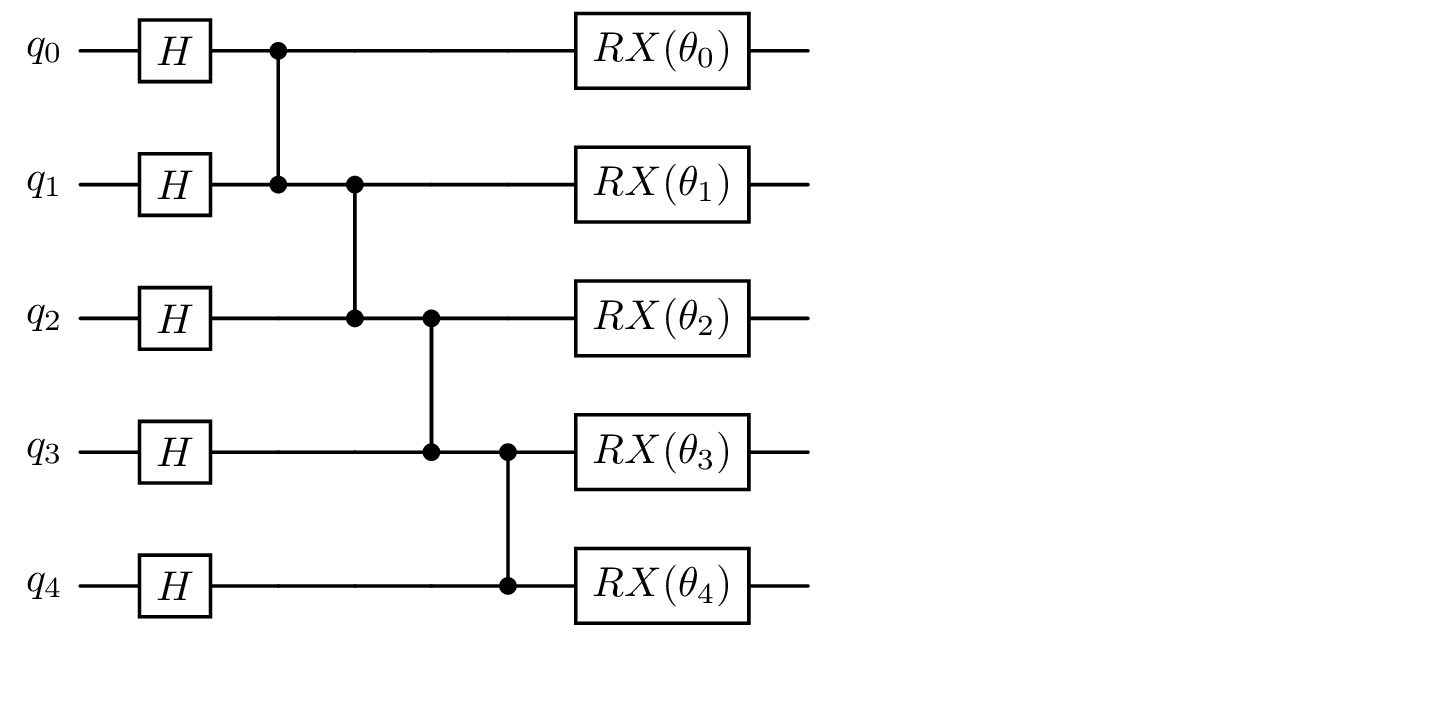}
		\caption{}
		\label{fig:Hadamard}
	\end{subfigure}
	\hfill
	\begin{subfigure}[t]{\textwidth}
		\centering
		\includegraphics[width=\textwidth]{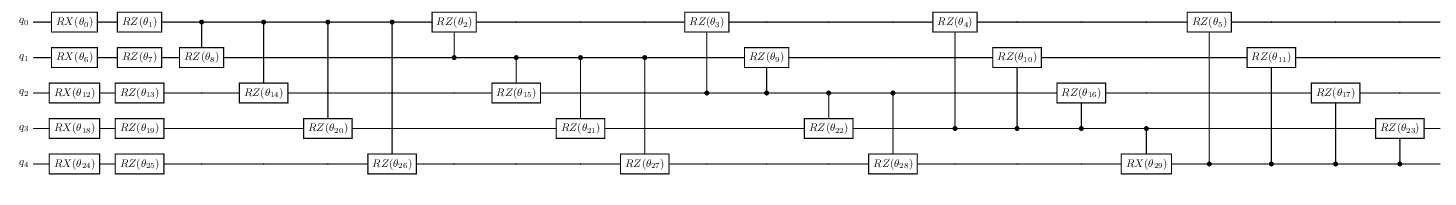}
		\caption{}
		\label{fig:Full-CRZ}
	\end{subfigure}
	\hfill
	\begin{subfigure}[t]{\textwidth}
		\centering
		\includegraphics[width=\textwidth]{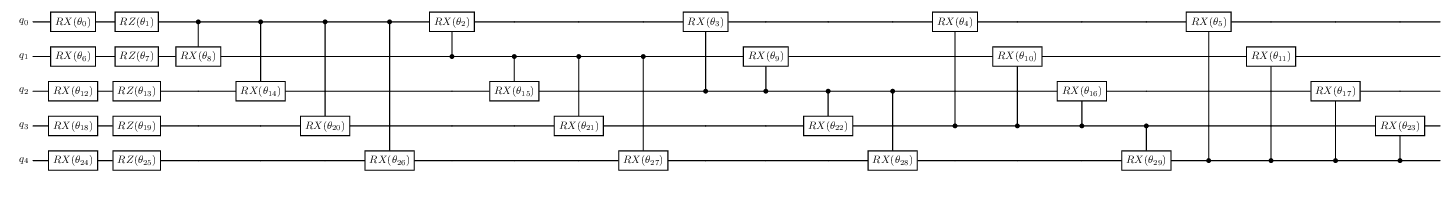}
		\caption{}
		\label{fig:Full-CRX}
	\end{subfigure}
	\caption{The 12 variational circuits examined herein include (a) Modified-Pauli-CRZ, (b) Modified-Pauli-CRX, (c) Efficient-CRZ, (d) Efficient-CRX, (e) HWE-CNOT, (f) HWE-CZ, (g) Full-Pauli-CRZ, (h) Full-Pauli-CRX, (i) ESU2, (j) Hadamard, (k) Full-CRZ, and (l) Full-CRX.}
	\label{fig:ansatz}
\end{figure}

The next layer in a PQC, that follows the encoding layer, is the variational layer.
These layers introduce trainable parameters into the quantum circuit, enabling optimization through classical computation and providing flexibility to the QML models.
These layers have a general form, defined as,
\begin{equation}
	U(\bm{\theta}) = \prod_{v} U_{v}(\theta_{v}), 
	\label{eq:general_variational}
\end{equation}
where $\bm{\theta}$ denotes the variational parameters, $v$ denotes the number of times that the layer is repeated within the circuit, and all entanglement gates are implicitly included in $U_{v}(\theta_{v})$.
In our study, we examine 12 variational circuits, found in \citet{sim_expressibility_2019}, as displayed in Fig.~\ref{fig:ansatz}.

To introduce more model parameters, increase the expressibility, and the nonlinearity of the PQCs, two expansion strategies can be explored.
The first we denoted as the number of ansatz layers (AL), where $v$ can be increased to introduce more unique trainable parameters into the PQC.
The second expansion strategy builds on the work of P\'{e}rez-Salinas \textit{et al.} that showed data re-uploading in PQCs is equivalent to the Universal Approximation Theorem for artificial neural networks.\cite{perez-salinas_data_2020}
The goal of data re-uploading is to introduce additional non-linearity in the model to help the PQC learn more complex functions, which is achieved by combining Eqs. \ref{eq:general_encoding} and \ref{eq:general_variational} into a general circuit,

\begin{equation}
	\ket{\Psi} =\prod_{k}\left( U(\bm{\theta}) U_{\Phi(\mathbf{x})}\right)\ket{0}^{\otimes n} = \prod_{k}
	\left( \prod_{v} U_{v}(\theta_{v}) \prod_{l} E_{\text{ent}}^{l} U_{\phi_{l}(\mathbf{x})} \right)  \ket{0}^{\otimes n},
\end{equation}
where $k$ denotes the re-upload depth (RUD) of the circuit.

The last component of a PQC is the measurement, which is required to recover the $i$th-predicted target value, $\hat{y}_{i}$, of the machine learning model that is used to update the model parameters.
This is performed by measuring the quantum state, $\ket{\Psi}$, using the Pauli Z operator on the first qubit denoted as,
\begin{equation}
	\hat{y}_{i} = \bra{\Psi}Z_{0}\ket{\Psi}_{i},
	\label{eq:y_pred}
\end{equation}
and passing the set of predicted target values, $\bm{\hat{y}} = (\hat{y}_{1},\ldots,\hat{y}_{i}, \ldots, \hat{y}_{N}) \in \mathbb{R}^{N}$, where $N$ is the number of samples, to the loss function, $\mathcal{L}(\bm{y}, \bm{\hat y})$, where $y_{i}$ belongs to the set of true target values $\bm{y} = (y_{1}, \ldots, y_{i},\ldots, y_{N}) \in \mathbb{R}^{N}$.
While $\mathcal{L}$ can be any loss function relevant for regression-based ML tasks, here, we choose to use mean square error as the loss function,
\begin{equation}
	\mathcal{L}(\bm{y}, \bm{\hat y}) = \frac{1}{N} \sum_{i=1}^{N} (y_{i} - \hat{y}_{i})^{2}.
	\label{eq:isthisloss}
\end{equation}

\subsection{Computational Details}\label{subsection:CompDet}
To explore the large combination of PQCs previously mentioned, we introduce \textsc{qregress}, a modular Python package based on \textsc{PennyLane}\cite{bergholm_pennylane_2022} and \textsc{Qiskit}\cite{javadi-abhari_quantum_2024}.
The portion of the code based on \textsc{PennyLane} is capable of performing state-vector simulation using \textsc{Qulacs}\cite{suzuki_qulacs_2021}, noisy simulation using \textit{qiskit-aer} with the \textit{FakeQuebec} backend, along with access to real devices on the IBM Quantum Platform using the \textsc{PennyLane-Qiskit} plugin.
Additionally, to train and test regression models that require many circuit executions, we ported our \textsc{PennyLane} code to \textsc{Qiskit} so that the \textsc{Qiskit} Batch Execution mode could be utilized.
This was a vital step for running on \textit{ibm\_quebec} since \textsc{Qiskit} Sessions, as implemented in the \textsc{PennyLane-Qiskit} plugin, are too cumbersome to train PQCs with a large number of training samples.
In practice, all state-vector simulations were performed using \textsc{PennyLane}, except those that are used for direct comparisons with the \textit{FakeQuebec} and \textit{ibm\_quebec} backends, which were performed using the \textsc{Qiskit} code.
All state-vector simulations with 5 qubits were optimized using 1000 iterations, while 16 qubit simulations were initially run using 250 iterations due to the increased computational resources required to run these experiments.
State-vector simulations ran using \textsc{PennyLane} were optimized using the Simultaneous Perturbation Stochastic Approximation method (SPSA) since we found that SPSA requires fewer steps to optimize when compared to other optimizers implemented in \textsc{PennyLane}.
For the experiments run using the \textsc{Qiskit} code, we utilized the constrained optimization by linear approximation (COBYLA) optimizer, as implemented in \textsc{SciPy}\cite{virtanen_scipy_2020}.
For all models, features ($\mathbf{x}$) and target values ($\mathbf{y}$) were scaled using the MinMaxScaler, as implemented in \textsc{Scikit-learn}\cite{pedregosa_scikit-learn_2011}, such that all features and target values are $\mathbb{R}\in [ -1,1 ]$.


\subsection{Datasets}
Herein, we use two datasets based on quantum chemistry that provide unique insights into the behavior of PQCs.
The first dataset we analyze is the BSE49 dataset\cite{prasad_bse49_2021}, which is useful for benchmarking ML-based applications. 
This approach aligns with workflows familiar to most computational chemists using machine learning---i.e., taking the Cartesian coordinates of a molecule, transforming them into input data using a classical molecular representation, and predicting a chemically relevant property.
The second dataset we explore, based on the DDCC method of \citet{townsend_data-driven_2019}, is more domain-specific since both the input and output of the machine learning models are based on electronic structure data, as opposed to a structural molecular representation, such as Molecular ACCess Systems (MACCS)\cite{durant_reoptimization_2002} or Morgan fingerprints \cite{morgan_generation_1965,rogers_extended-connectivity_2010}.
Due to the formulation of both datasets, they offer two unique challenges for the PQCs we study, which are highlighted in the results and discussion section below.

\begin{figure}[H]
	\centering
	\begin{subfigure}[b]{\textwidth}
		\centering
		\includegraphics[width=\textwidth]{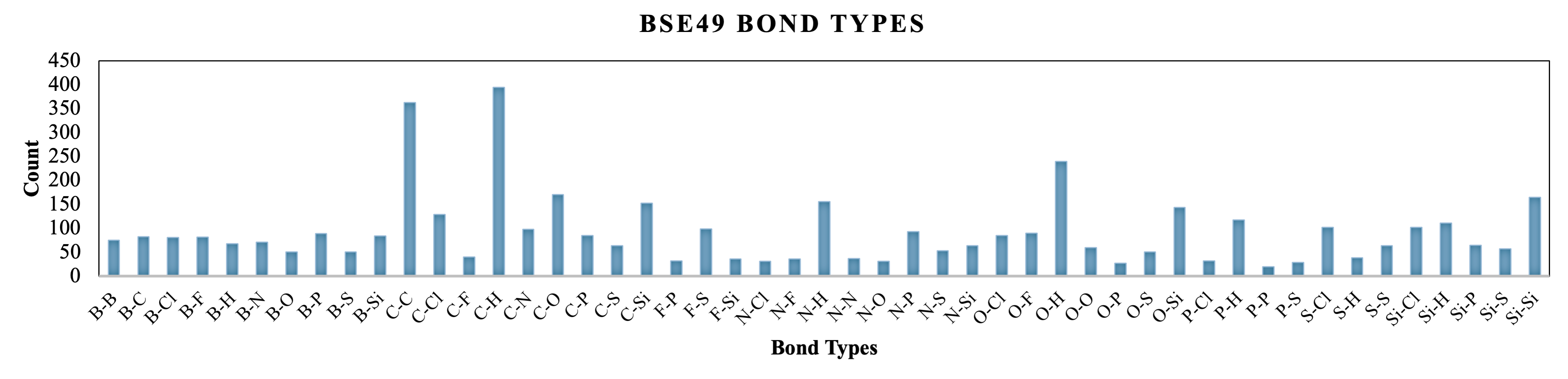}
		\caption{}
		\label{fig:bondtypes}
	\end{subfigure}
	\hfill
	\begin{subfigure}[b]{0.49\textwidth}
		\centering
		\includegraphics[width=\textwidth]{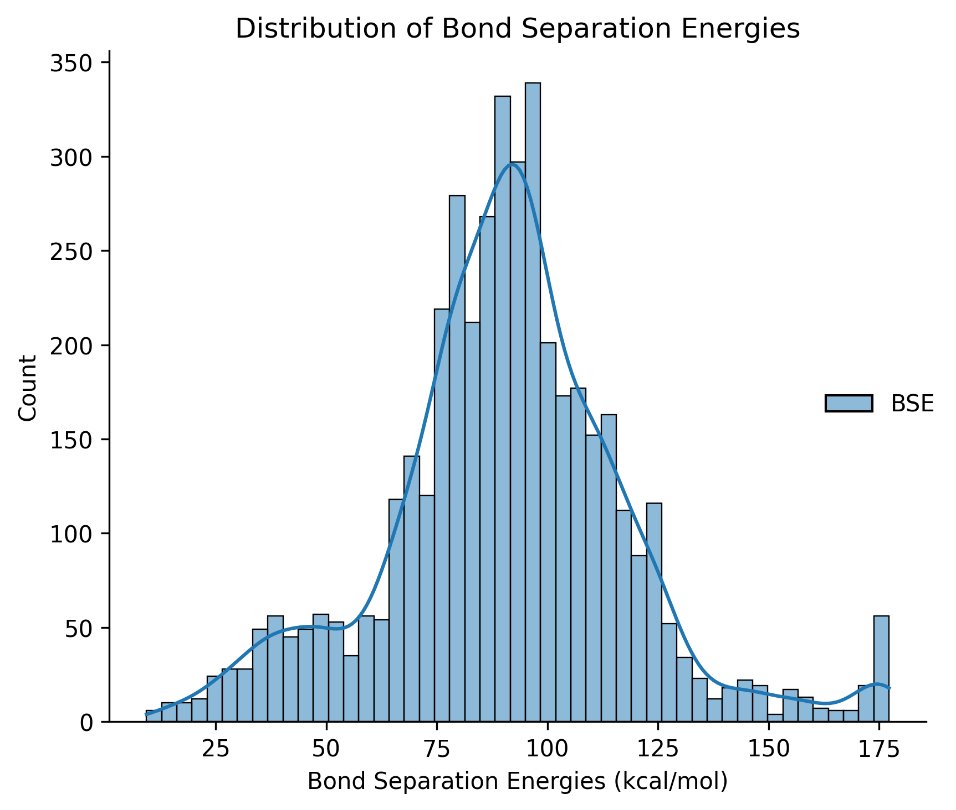}
		\caption{}
		\label{fig:BSEdistr}
	\end{subfigure}
	\hfill
	\begin{subfigure}[b]{0.49\textwidth}
		\centering
		\includegraphics[width=\textwidth]{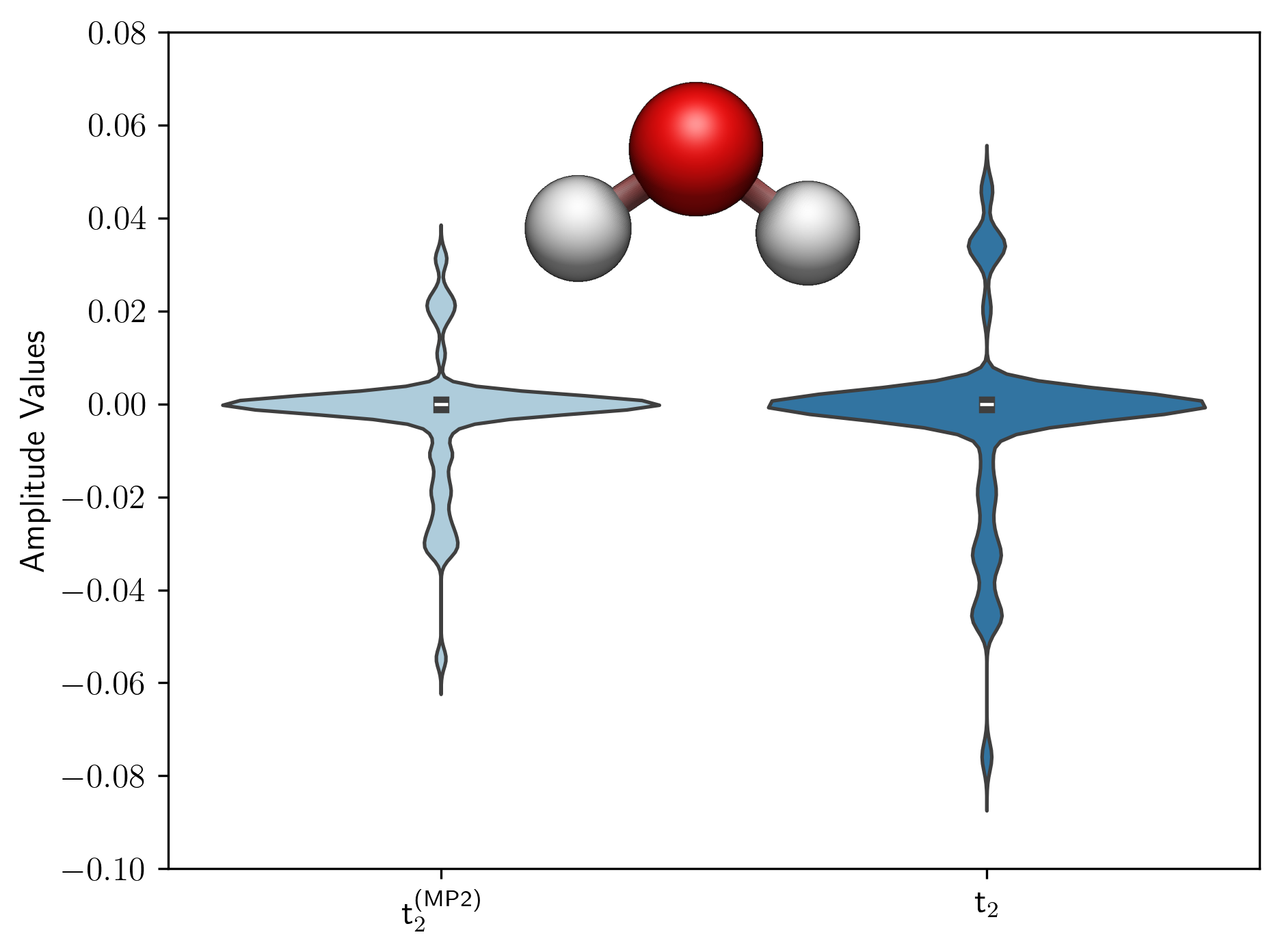}
		\caption{}
		\label{fig:waterddccdistribution}
	\end{subfigure}	
	\caption{Distributions of the (a) bond types and the (b) bond separation energies (in kcal/mol) for the BSE49 dataset, along with the distributions of the (c) initial MP2 $t_{2}$-amplitudes (left) and the optimized CCSD $t_{2}$-amplitudes (right).}
	\label{fig:bse_data}
\end{figure}

The BSE49 dataset contains bond separation energies (BSEs) for homolytic bond cleavage of covalently bonded molecules, such as \ce{A-B -> A^{.} + B^{.}}.\cite{prasad_bse49_2021}
This dataset consists of 49 unique A-B bonds (Fig.~\ref{fig:bondtypes}) with a large spread of bond separation energies (BSEs), ranging from 9.38 to 177.24 kcal/mol, as highlighted in Fig.~\ref{fig:BSEdistr}.
Overall, there are 4394 data points, 1951 of which are existing structures, while the remaining 2443 are hypothetical structures, calculated using the (RO)CBS-QB3\cite{wood_restricted-open-shell_2006,montgomery_complete_1999,montgomery_complete_2000} composite quantum chemistry method.
Like many datasets composed of molecular structures, often in Cartesian (XYZ) format, an important aspect of the data preprocessing is the choice of molecular representation, or how the molecule is represented in the machine learning models.\cite{jones_molecular_2023}
In general, molecular representations can be partitioned into three groups, graph-, topology-, and physics-based representations, all of which were explored to provide a comprehensive overview of how they perform using classical models as a baseline, as highlighted in Supplementary Information (SI) Section \ref{section:BSE49_Feature_Set}.
During the preprocessing stage, before converting the XYZ coordinates into our molecular representation of choice, the set of hypothetical structures was reduced to 2436 molecules due to issues with valency exceptions when converting XYZ coordinates into \textsc{RDKit} mol objects.
Using \textsc{RDKit}\cite{noauthor_rdkit_nodate}, we examined three commonly applied graph-based molecular representations, Molecular ACCess Systems (MACCS)\cite{durant_reoptimization_2002}, Morgan or extended-connectivity fingerprints \cite{morgan_generation_1965,rogers_extended-connectivity_2010}, and \textsc{RDKit} fingerprints.
All three of these methods use traversals of the molecular graphs to encode various structural details into bit vectors.
Lastly, we explore both topology- and physics-based molecular representations, both of which encode the three-dimensional structure of molecules in unique ways.
Persistent images (PIs) are a topology-based fingerprint that uses persistence homology to encode topological information of three-dimensional molecular structures into fixed-dimension images.\cite{adams_persistence_2017,townsend_representation_2020,schiff_augmenting_2022} 
We use the implementation from Townsend \textit{et al.}\cite{townsend_representation_2020}, which uses the \textsc{Ripser} Python package to generate PIs.\cite{tralie_ripserpy_2018}
Lastly, we explore two physics-based representations, Coulomb matrices (CMs) \cite{rupp_fast_2012} and smooth overlap of atomic positions (SOAPs), that were generated using \textsc{DScribe}.\cite{de_comparing_2016}
Both methods encode physical information regarding the atomic environments of each molecule, where CMs encode the Coulomb repulsion between atoms using nuclear charges, while SOAPs encode the similarities between atomic neighborhoods using kernels.
Due to the computational cost associated with computing the regularized entropy match (REMatch) kernel over the set of 2436 molecules with the SOAPs representation, we excluded SOAPs from the overall discussion.
For each of the molecular representations, we analyzed two different methods for representing the components of the BSE reaction.
The first represents the feature vector as the products subtracted from the reactants, denoted as \textit{sub} (e.g., $\mathbf{X}_{\mathit{sub}} = (\mathbf{X}_{\text{A}^{.}} + \mathbf{X}_{\text{B}^{.}}) - \mathbf{X}_{\text{A-B}}$), similar to the method used in the study of \citet{garcia-andrade_barrier_2023}, and the second represents the feature vector using only the reactant (A-B), denoted as \textit{AB} (e.g., $\mathbf{X}_{\mathit{AB}} =  \mathbf{X}_{\text{A-B}}$).
Based on the results demonstrated in Figs. \ref{fig:classical_molrepfig} and \ref{fig:bse_classical_features}, experiments using the BSE49 data are reduced to 5 or 16 qubits using principal component analysis (PCA) applied to the Morgan fingerprints using the \textit{sub} formulation.

The second quantum chemistry dataset we analyze in this paper is generated using the data-driven coupled-cluster (DDCC) approach of Townsend and Vogiatzis.\cite{townsend_data-driven_2019,jones_chapter_2023}
DDCC is an ML-based approach for accelerating the convergence of coupled-cluster singles and doubles (CCSD) calculations by predicting the $t_{2}$-amplitudes of the CCSD wave function (SI Eq. \siref{eq:cc_wfn}) with features generated using lower-level methods that are used to initialize CCSD calculations, such as Hartree-Fock (HF) and M{\o}ller-Plesset second-order perturbation theory (MP2).
In SI Section \ref{section:DDCC_Feature_Set}, we provide a brief overview of the equations required to understand the DDCC method.
The DDCC feature set consists of the MP2 $t_{2}$-amplitudes used to initialized CCSD, along with two-electron integrals ($\mel{ij}{}{ab}$), the difference between the occupied and virtual orbital energies ($\varepsilon_{i}+\varepsilon_{j}-\varepsilon_{a}-\varepsilon_{b}$), a binary feature to denote whether the excitation goes to the same virtual orbital ($a=b$), and the orbital energies ($\varepsilon_{i},\varepsilon_{j},\varepsilon_{a},\varepsilon_{b}$).
Additionally, the feature set includes terms related to the individual contributions to the orbital energies, such as the one-electron Hamiltonian ($h$), Coulomb matrix ($J$), exchange matrix ($K$), and Coulomb and exchange integrals ($J^{i}_{a}, J^{j}_{b}, K^{a}_{i}, K^{b}_{j}$).
In total, there are 30 features for each CCSD $t_{2}$-amplitude due to the addition of features that denote the sign and magnitude of the MP2 $t_{2}$-amplitudes.

The dataset analyzed in this study consists of 199 water conformers obtained from the study by Townsend and Vogiatzis\cite{townsend_data-driven_2019}, generated using the STO-3G basis set\cite{hehre_selfconsistent_1970} with the frozen core approximation, as implemented in \textsc{Psi4}\cite{parrish_psi4_2017} and \textsc{\textsc{Psi4}Numpy}\cite{smith_psi4numpy_2018}.
Like with the BSE49 dataset, we reduce the number of features to 5 or 16 qubits.
As emphasized in Fig.~\ref{fig:DDCC_feature_set}, when SHapley Additive ExPlanation analysis (SHAP)\cite{lundberg_unified_2017} is applied to the DDCC feature set, 5 features, which includes the two-electron integrals ($\mel{ij}{}{ab}$), MP2 $t_{2}$-amplitudes ($t^{ab}_{ij(\text{MP2})}$), the magnitude of the MP2 $t_{2}$-amplitudes, the difference in orbital energies ($\varepsilon_{i}+\varepsilon_{j}-\varepsilon_{a}-\varepsilon_{b}$), and the binary feature denoting whether two-electrons are promoted to the same virtual orbital ($a=b$), are sufficient to achieve the same accuracy as the set consisting of 30 features.
We will also note that despite using a minimal basis set, this is a data-intensive quantum chemistry application for PQCs since for each molecule there are $(N_{occ})^{2}(N_{virt})^{2}$ $t_{2}$-amplitudes, where $N_{occ}$ denotes the number of occupied orbitals and $N_{virt}$ denotes the number of virtual orbitals.
For water conformers with 4 occupied and 2 virtual orbitals, there are a total of 64 $t_{2}$-amplitudes per molecule; therefore, for 199 water molecules, there are 12,736 $t_{2}$-amplitudes, where the distributions of the initial MP2 and calculated-CCSD $t_{2}$-amplitudes are highlighted in Fig.~\ref{fig:waterddccdistribution}.

\section{Results}

\begin{figure}[H]
    \centering	
    \begin{subfigure}[b]{\textwidth}
        \centering
        \includegraphics[width=\linewidth]{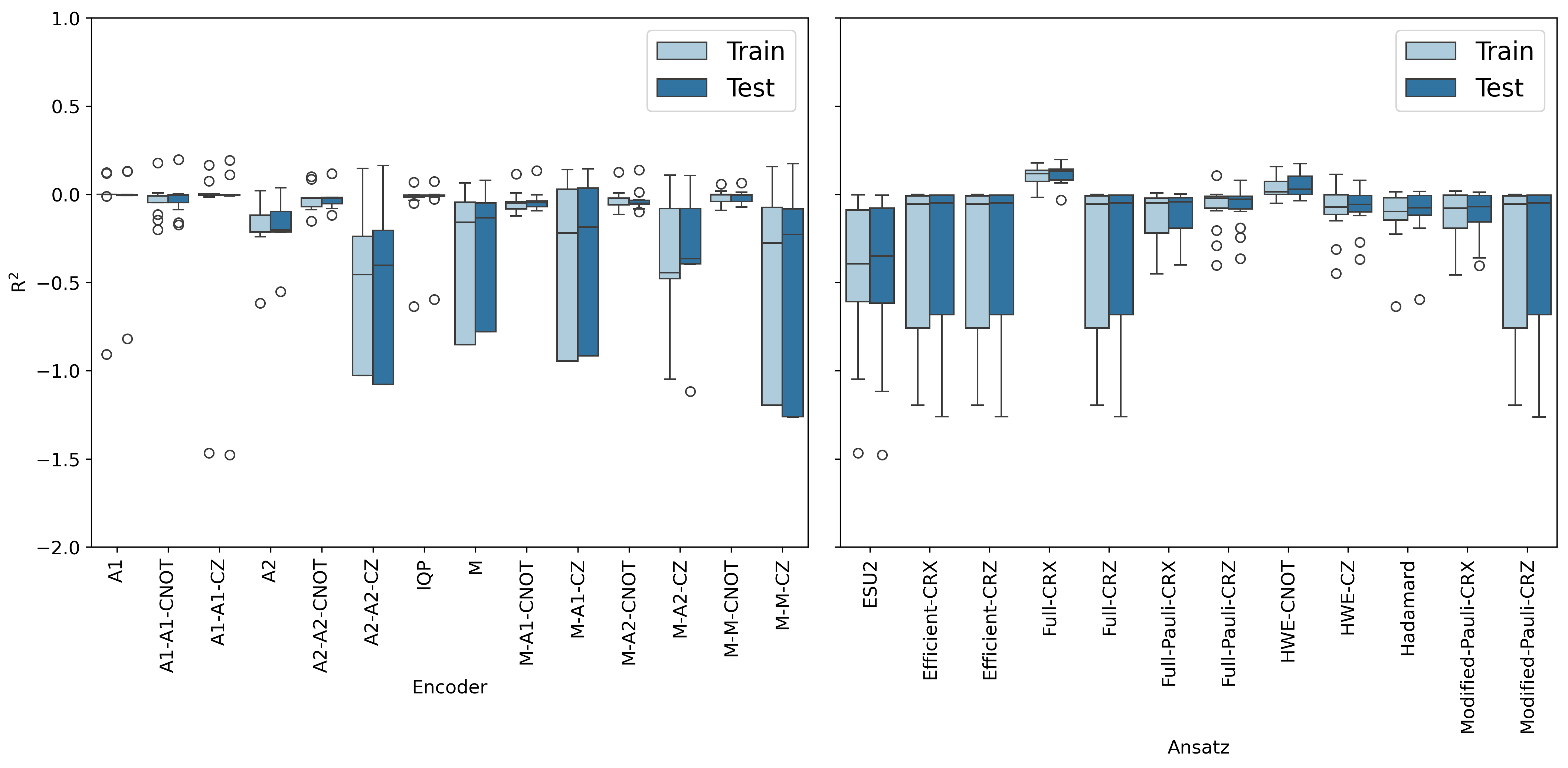}
        \caption{}
        \label{fig:5BSE_boxplots}
    \end{subfigure}
    \hfill
    \begin{subfigure}[b]{\textwidth}
        \centering
        \includegraphics[width=\linewidth]{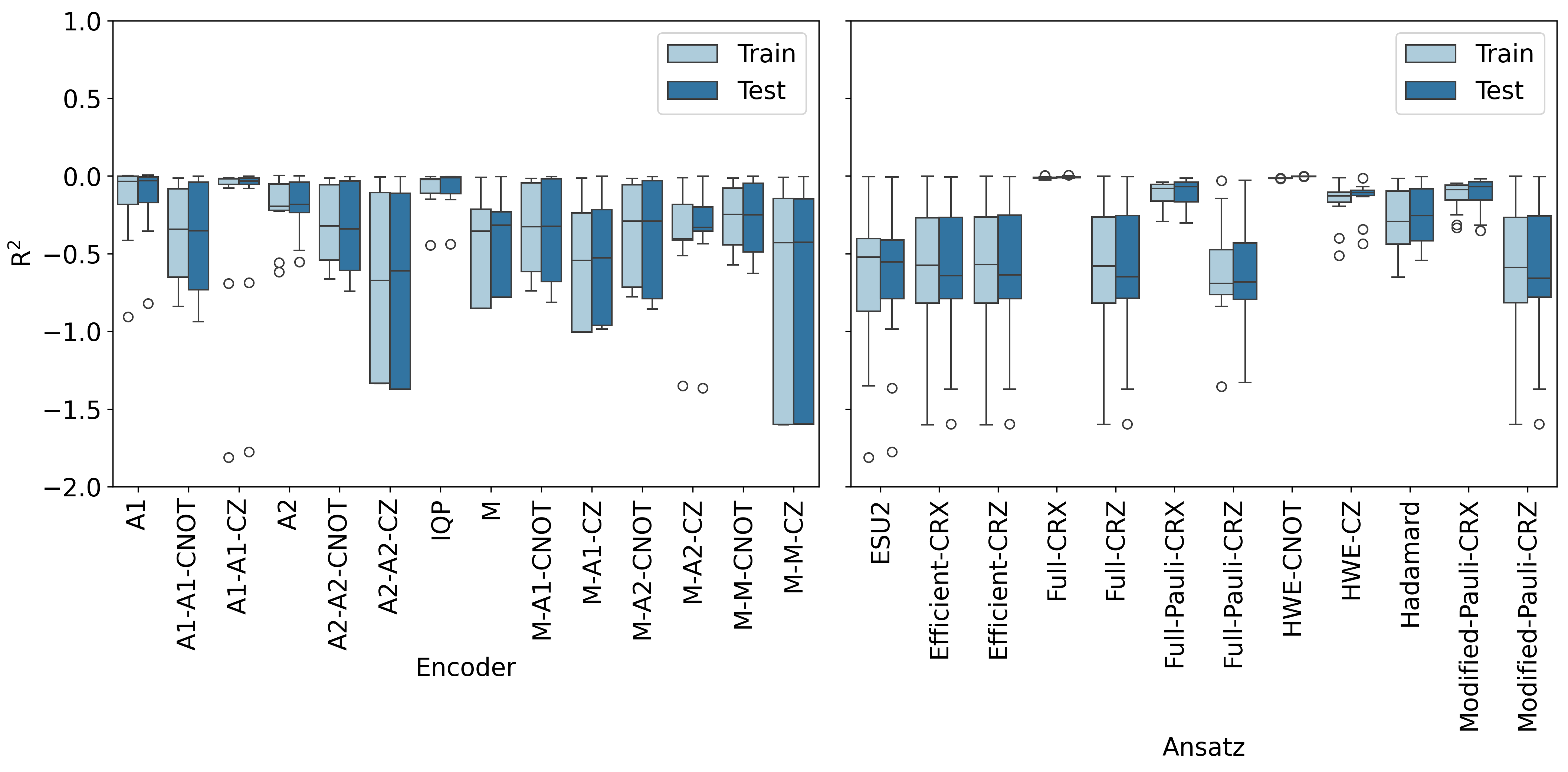}
        \caption{}
        \label{fig:16BSE_boxplots}
    \end{subfigure}
    \caption{The model performance (in R$^{2}$) of the 168 parametrized quantum circuits for BSE49 using box plots to highlight the general performance of the encoder (left) and ansatz (right) circuits for the (a) 5 and (b) 16 qubit BSE49 models. Additionally, outliers in the data are denoted using circles with a black outline.}
    \label{fig:BSEboxandheat}	
\end{figure}

In this section, we analyze the unique set of 168 parameterized quantum circuits (PQCs) to identify the most suitable encoder–ansatz pair for each dataset.
Due to the cost of state-vector simulation, a re-upload depth of 1 ($k=1$) and a single ansatz layer ($v=1$) are used to obtain insights into this broad set of circuits.
Starting with the 5-qubit BSE49 dataset, Fig.~\ref{fig:5BSE_heatplots} presents a heatmap of the performance of all 168 PQCs, displaying the coefficients of determination (R$^{2}$) for both the training set (left) and test set (right).
The training set R$^2$ values range from a minimum of $-1.4673$ to a maximum of $0.1775$, with a mean of $-0.1865$ and a standard deviation of $0.3433$. 
For the test set, the R$^2$ values range from a minimum of -$1.4774$ to a maximum of $0.1974$, with a mean of $-0.1765$ and a standard deviation of $0.3446$.
General trends regarding the performance of the encoders (left) and ansatz (right) layers are highlighted in Fig.~\ref{fig:5BSE_boxplots} using box plots.
On average, the best encoding layer is M-M-CNOT with training and test R$^{2}$ values of $-0.0173$ and $-0.0157$, respectively.
The best ansatz layer is Full-CRX with an average training  R$^{2}$ of $0.1054$ and test R$^{2}$ of $0.1165$.
Overall, for the BSE49 dataset using 5 qubits, the best encoder-ansatz pair is A1-A1-CNOT{\_}Full-CRX which has a training set R$^{2}$ of $0.1775$ and test set R$^{2}$ of $0.1974$.
Although A1-A1-CNOT{\_}Full-CRX was the best-performing encoder–ansatz pair, we selected M-M-CZ{\_}HWE-CNOT for our circuit depth exploration. This pair achieves comparable performance, with an R$^{2}$ of $0.1566$ on the training set and $0.1740$ on the test set, but results in a shallower circuit, making it more computationally efficient for state-vector simulations.

Following the 5 qubit models, we examined the BSE49 dataset on the set of 168 PQCs using 16 qubits.
Fig.~\ref{fig:16BSE_heatplots} shows the performance in R$^{2}$ for the training (left) and test (right) sets, where the best encoder-ansatz pair is A1{\_}Efficient-CRZ with an R$^{2}$ of $-0.0007$ for the training set and $-0.0050$ for the test set.
Across the 168 PQCs, the training and test sets have a minimum R$^{2}$ of $-1.8104$ and $-1.7752$, mean of $-0.4045$ and $-0.4022$, maximum of $-0.0007$ and $-0.0004$, and standard deviation of $0.4161$ and $0.4235$, respectively.
Overall, as highlighted in Fig.~\ref{fig:16BSE_boxplots} (left), the best encoder is IQP, with a mean R$^{2}$ of $-0.0942$ for the training set and $-0.0882$ for the test set.
The best ansatz circuit (Fig.~\ref{fig:16BSE_boxplots} right) is HWE-CNOT with a mean R$^{2}$ of $-0.0148$ and $-0.0025$ for the training and test set, respectively.
Due to the overall lower performance and the high computational cost of state-vector simulations with 16 qubits, we limited our additional experiments on the BSE49 dataset to the best-performing 5-qubit PQC, M-M-CZ{\_}HWE-CNOT.

\begin{figure}[H]
    \centering	
    \includegraphics[width=\linewidth]{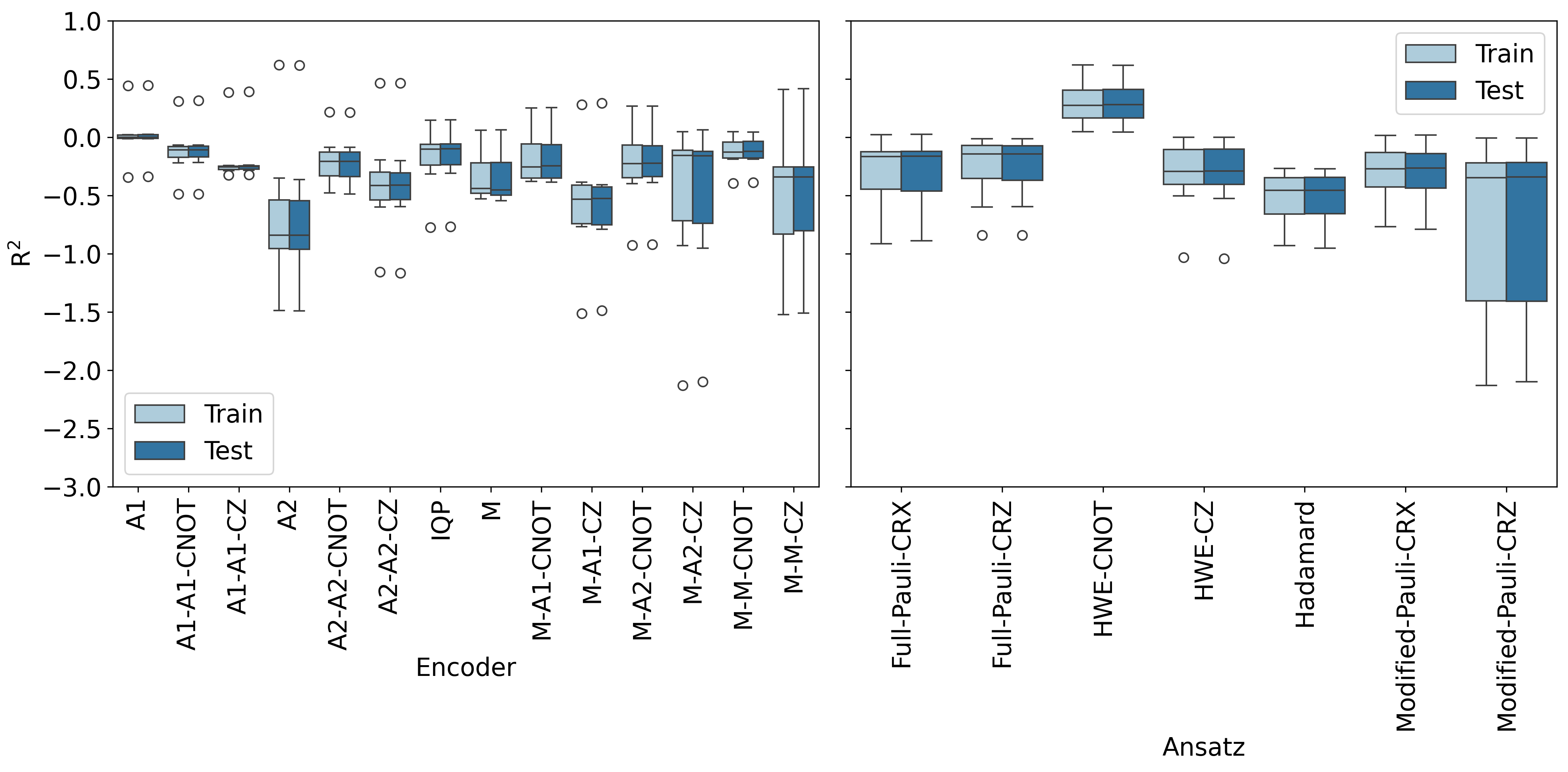}
    \caption{The model performance, using the coefficient of determination (R$^{2}$), for the reduced set of 98 PQCs using the DDCC dataset, for the encoders (left) and variational (right) layers is highlighted in (b).}	
    \label{fig:ddccboxplots}
\end{figure}

Following the initial analysis of the 168 PQCs using the BSE49 dataset, we reduced the set to 98 PQCs for the DDCC dataset based on circuit depth and accuracy.
Fig.~\ref{fig:ddccheatplots} highlights the improved performance of the PQCs on the DDCC dataset over the BSE49 dataset, where the R$^{2}$ values of the training set have a minimum, mean, maximum, and standard deviation of $-2.1302$, $-0.3050$, $0.6191$, and $0.4422$, respectively.
While the R$^{2}$ values of the test set have a minimum of $-2.1007$, a mean of $-0.3046$, a maximum of $0.6184$, and a standard deviation of $0.4411$.
The best encoder-ansatz pair is A2{\_}HWE-CNOT with R$^{2}$ values of $0.6191$ and $0.6184$ for the training and test sets, respectively.
We found that the PQCs offer better model performance for the DDCC dataset, with A1 offering the best average performance of the encoders with an R$^{2}$ of $0.0163$ for the training set and $0.0183$ for the test set (Fig.~\ref{fig:ddccboxplots} left).
Like the 16 qubit BSE49 data, the best ansatz layer is HWE-CNOT (Fig.~\ref{fig:ddccboxplots} right).
Unlike the 16 qubit BSE49 data, HWE-CNOT offers a training and test set R$^{2}$ of $0.2817$ and $0.2853$, respectively.
Overall, the DDCC models offer more accurate models, with less overfitting, when compared to the BSE49 models.

\begin{figure}[H]
	\centering	
	\begin{subfigure}[b]{\textwidth}
		\centering
		\includegraphics[width=\linewidth]{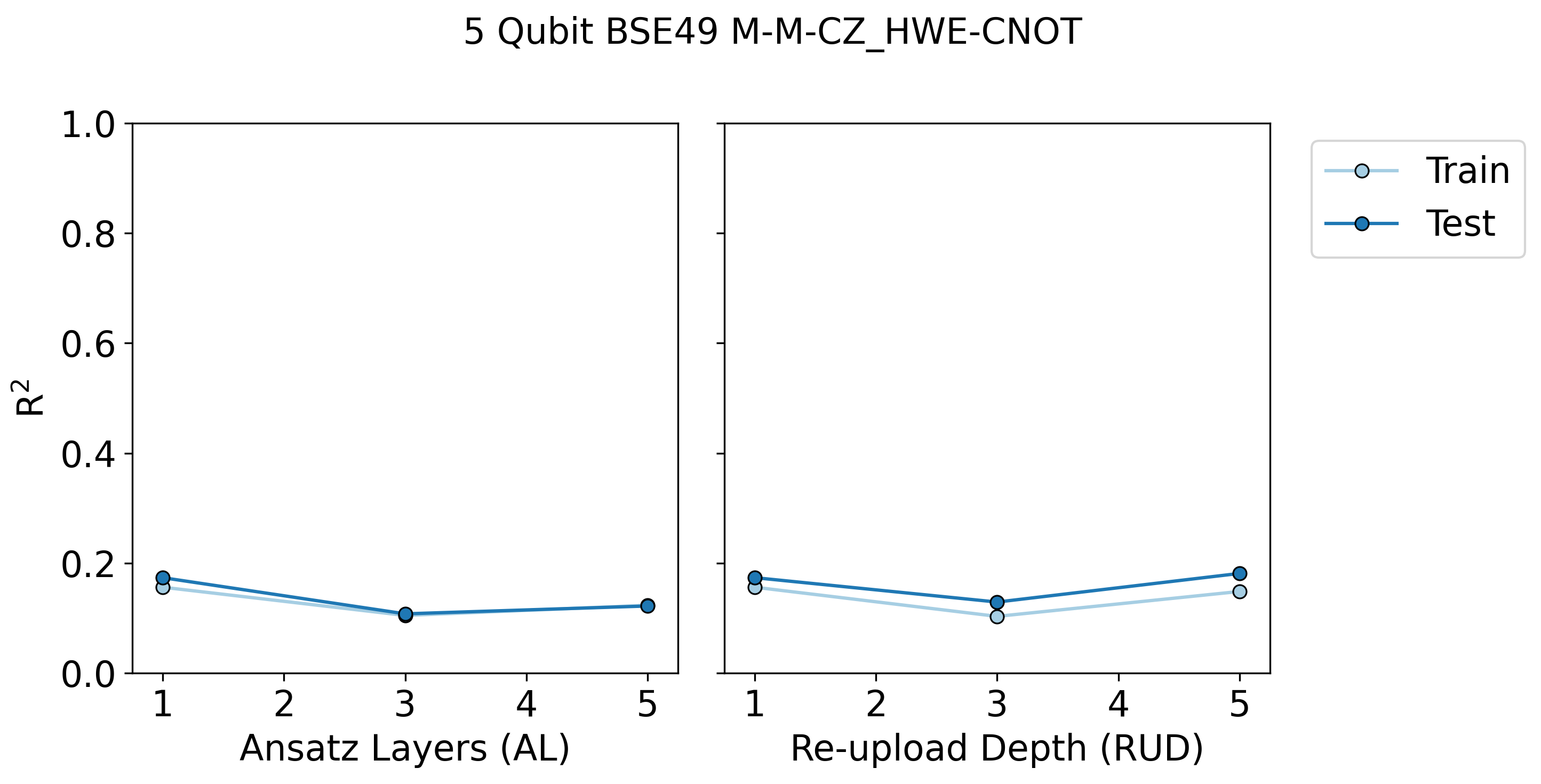}
		\caption{}
		\label{fig:bse5RUDAL_lineplot}
	\end{subfigure}
	\hfill
	\begin{subfigure}[b]{\textwidth}
		\centering
		\includegraphics[width=\linewidth]{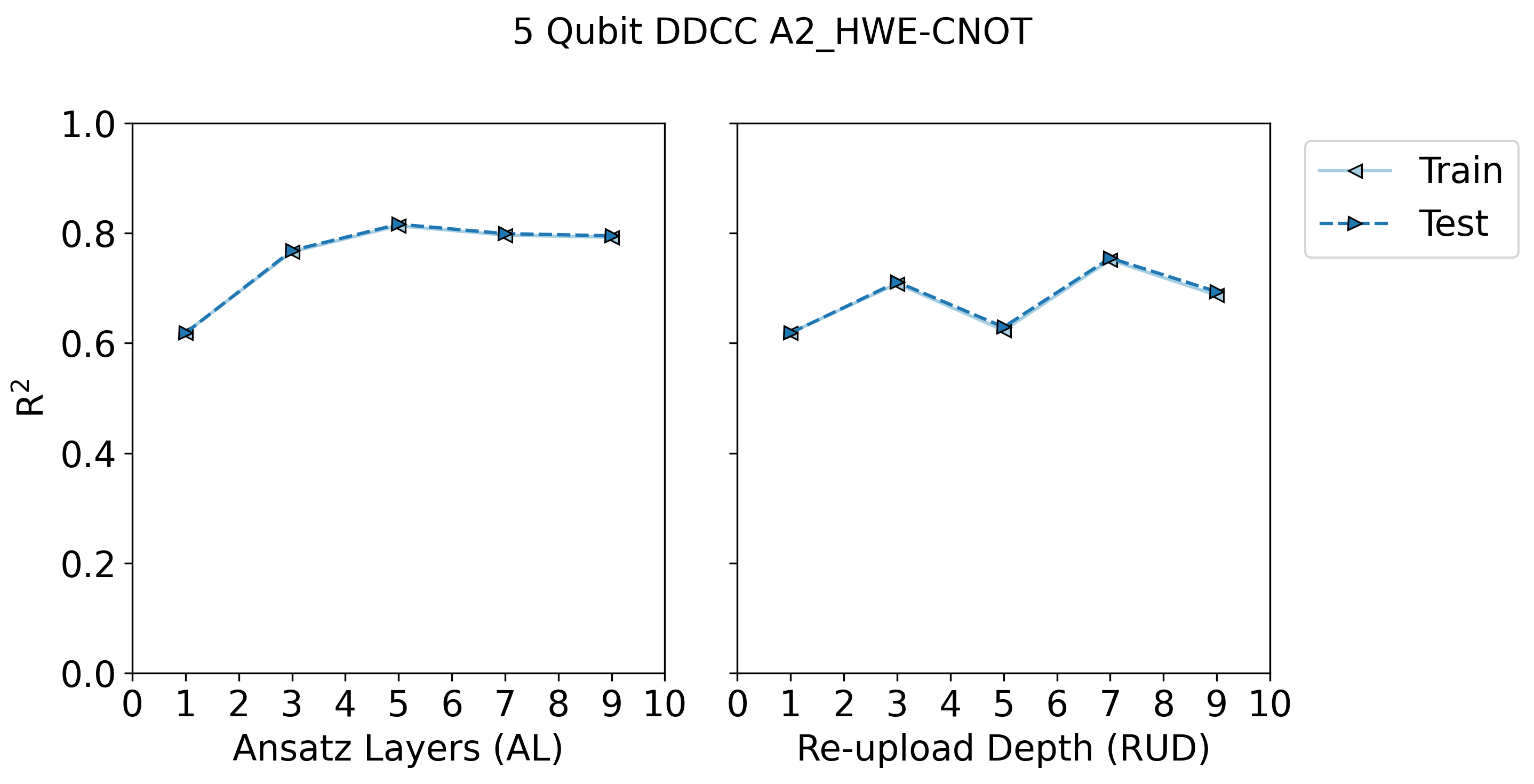}
		\caption{}
		\label{fig:ddccRUDAL_lineplot}
	\end{subfigure}

	\caption{Line plots to highlight the performance as the number of ansatz layers (AL; left) and re-upload depth (RUD; right) increases for the 5 qubit (a) BSE49 and (b) DDCC data.}
	\label{fig:RudAlLinePlot}	
\end{figure}

Following the evaluation of the initial set of PQCs on the BSE49 and DDCC datasets, we investigate how increasing the re-upload depth (RUD) and the number of ansatz layers (AL) impacts model performance.
The purpose of this investigation is to analyze reported claims that data re-uploading increases the model performance and that more model parameters (e.g., increasing the number of AL) increase model accuracy.\cite{perez-salinas_data_2020,suzuki_predicting_2020}
For the BSE49 data, we examine RUDs of $k=\{1,3,5\}$ and for the ALs $v=\{1,3,5\}$.
We examine this data in two ways, the first is a simple line plot, shown in Fig.~\ref{fig:bse5RUDAL_lineplot}, where the number of ALs and the RUD layers are plotted on the x-axis, the y-axis denotes the model accuracy in R$^{2}$, and the training set is denoted in light blue, while the test set is shown in dark blue.
The second is by analyzing the regression parity plots, Fig.~\ref{fig:BSE5_distribution_parity}, with the addition of kernel density estimation (KDE) plots on the axes to highlight the distributions of the target values.
Overall, for M-M-CZ{\_}HWE-CNOT using the BSE49 dataset, increasing the number of ALs and RUD offers negligible improvements over the base model, which has an R$^{2}$ of $0.16$ and $0.17$ for the training and test set, respectively.
The model with the best R$^{2}$ for the training set is the initial model using an AL of 1 and RUD 1, while the model with the best R$^{2}$ for the test set uses an AL of 1 and RUD of 5 where the R$^{2}$ is $0.18$.
When we analyze the distributions in Fig.~\ref{fig:BSE5_distribution_parity}, we find that the reference target values have a wide distribution, ranging from $9.38$ to $177.24$ kcal/mol with a standard deviation of $27.4877$ kcal/mol.
For the predicted values, all of the models struggle with predicting an accurate range, where the model with the worst spread is AL=3/RUD=1 with a minimum of $67.6736$ kcal/mol, a maximum of $115.6791$ kcal/mol, and a standard deviation of $6.7346$ kcal/mol.
The best model regarding the spread of the predicted data has an AL of 1 and RUD of 5, with a minimum of $49.8978$ kcal/mol, a maximum of $127.8078$ kcal/mol, and a standard deviation of $14.0201$ kcal/mol. 
Overall, all models, regardless of the number of ALs and RUD, tend to regress towards the mean value of the reference data, $90.6916$ kcal/mol.

When compared to the 5 qubit BSE49 data, the DDCC dataset using A2{\_}HWE-CNOT offers improved performance as we increase the number of ALs and RUD, as highlighed in Figs. \ref{fig:ddccRUDAL_lineplot} and \ref{fig:ddccdistribution_parity}.
Unlike the BSE49 model, we go beyond ALs and RUDs of 5 due to the initial improvements shown as both values increased.
As highlighted in Fig.~\ref{fig:ddccRUDAL_lineplot}, as the number of ALs increases, the model performance improves from a training and test R$^{2}$ of $0.62$ using an AL=1/RUD=1, to R$^{2}$ values of $0.81$ and $0.82$ for the training and test sets, respectively, when AL=5/RUD=1.
When the AL is increased beyond 5, we found that both the training and test set R$^{2}$ values plateau around $0.80$, implying that the model has hit capacity regarding the number of training parameters.
Concerning the RUDs, we found that the models offer slight improvements as the RUD increases ($k=7>k=3>k=1$), while also containing periodic dips in model performance ($k=9>k=5\ge k=1$).
Overall, the best model for A2{\_}HWE-CNOT with the DDCC dataset is the model with an AL of 5 and RUD of 1.
This is highlighted in Fig.~\ref{fig:ddccdistribution_parity}, where we see that the distribution of the predicted target values is closer to the spread of the reference values than the initial model with AL=1/RUD=1.

\begin{figure}[H]
	\centering	
	\begin{subfigure}[b]{0.75\textwidth}
		\centering
		\includegraphics[width=\linewidth]{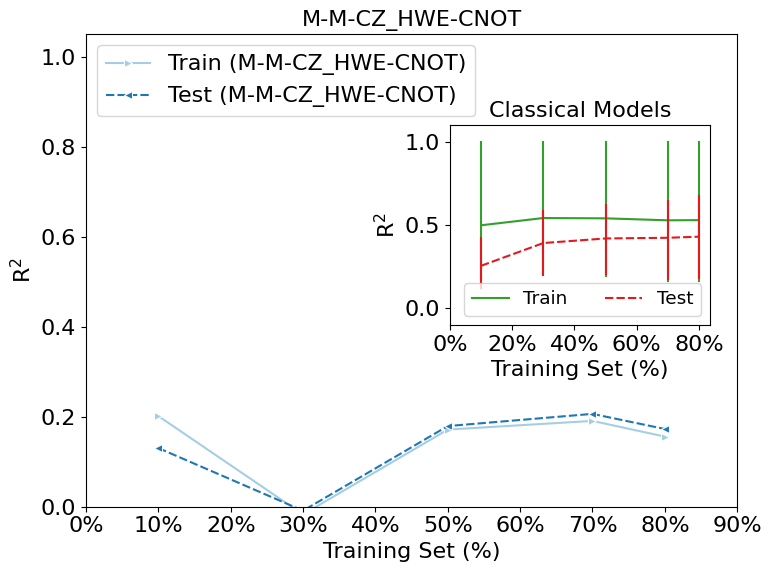}
		\caption{}
		\label{fig:BSE5_learning_curves}
	\end{subfigure}	
	\hfill
	\begin{subfigure}{0.75\textwidth}
		\centering
		\includegraphics[width=\linewidth]{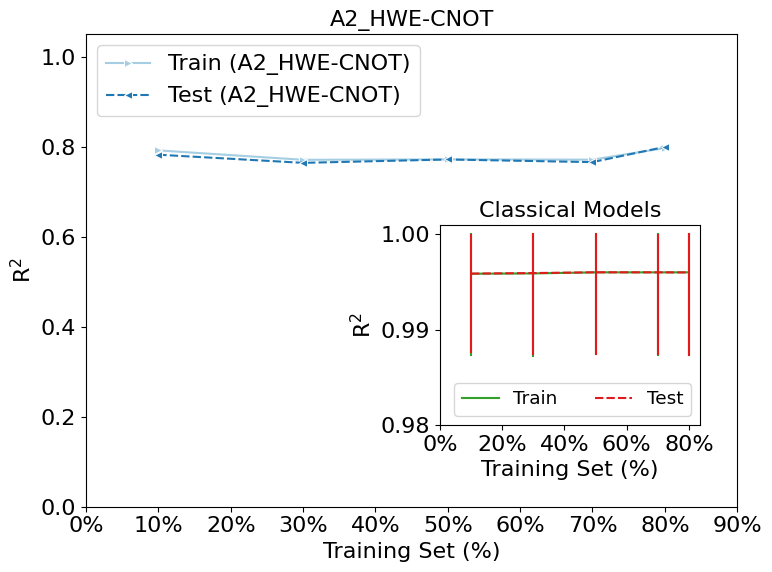}
		\caption{}
		\label{fig:ddcclearningcurves}
	\end{subfigure}
	\caption{Learning curves for the (a) BSE49 database using the M-M-CZ{\_}HWE-CNOT circuit and (b) the DDCC dataset using the A2{\_}HWE-CNOT circuit. The x-axis denotes the percentage of training points, while the test set is held to 20\% of the total data. The y-axis denotes the R$^{2}$ of the training and test learning curves. Inlay plots show the training and test data for the classical models examined, where the bottom error bar denotes the minimum value, the line is the mean value, and the top error bar is the maximum value.}
	\label{fig:learningcurves}
	
\end{figure}

Following the analysis of the RUD and number of ALs for both 5 qubit models, we explored claims that PQCs offer better model performance using fewer training points and hyperparameters\cite{schuld_circuit-centric_2020,du_expressive_2020,suzuki_predicting_2020} using learning curves.
We generate the learning curves by varying the number of training points from 10-80\% of the total data, while holding the number of test points to 20\% of the remaining dataset.
In Fig.~\ref{fig:learningcurves}, we analyze the effects of the number of training points of the best 5 qubits BSE49 and DDCC circuits, M-M-CZ{\_}HWE-CNOT using a RUD and AL of 1 and A2{\_}HWE-CNOT using a RUD of 1 and AL of 5.
For the BSE49 dataset, using 5 qubits, we found that M-M-CZ{\_}HWE-CNOT offers less overfit models but offers poorer performance when compared to the classical models examined, as highlighted in Fig.~\ref{fig:BSE5_learning_curves}.
Across the learning curve, the training set of the classical models has a mean R$^{2}$ of $0.5268$, a standard deviation of $0.3260$, a minimum R$^{2}$ of $0.1617$, and a maximum R$^{2}$ of $0.9999$.
Regarding the test set, the classical models are overfit, where the mean R$^{2}$ is $0.3832$, the standard deviation is $0.1762$, the minimum R$^{2}$ is $0.1226$, and the maximum R$^{2}$ is $0.6725$.
When we compare M-M-CZ{\_}HWE-CNOT with the classical models, we found that the training and test sets have mean R$^{2}$ values of $0.1405$ and $0.1360$, standard deviations of $0.0899$ and $0.0854$, minimum values of $-0.0172$ and $-0.0088$, and maximum values of $0.2015$ and $0.2064$, respectively.

When we perform the same analysis using the A2{\_}HWE-CNOT circuit on the DDCC dataset, we found that these models offer better performance than the BSE49 data, as highlighted in Fig.~\ref{fig:ddcclearningcurves}.
Like the BSE49 data, the classical models outperform A2{\_}HWE-CNOT across the learning curve with training and test mean R$^{2}$ values of $0.9960$ and $0.9960$, standard deviations of $0.0056$ and $0.0055$, minimum values of $0.9872$ and $0.9873$, and maximum values of $1.0000$ and $1.0000$, respectively.
For the training set, A2{\_}HWE-CNOT has a mean R$^{2}$ of $0.7808$, a standard deviation of $0.0128$, a minimum value of $0.7711$, and a maximum value of $0.7972$.
The test set offers similar performance with a mean R$^{2}$ of $0.7769$, a standard deviation of $0.0146$, a minimum value of $0.7643$, and a maximum value of $0.7997$.

For both circuits examined, two commonalities exist regarding the model performance.
The first is that both models are outperformed by their classical counterparts regarding both the wall clock time required to train and test the models and the accuracy of the models.
The second trend is more noticeable with the BSE49/M-M-CZ{\_}HWE-CNOT model, which offers fewer overfit models than the classical models, regardless of model accuracy.
While the second trend is less obvious for the DDCC/A2{\_}HWE-CNOT model, this model also lacks overfitting, which is characteristic of the classical methods.

\begin{figure}[H]
    \centering
    \includegraphics[width=\linewidth]{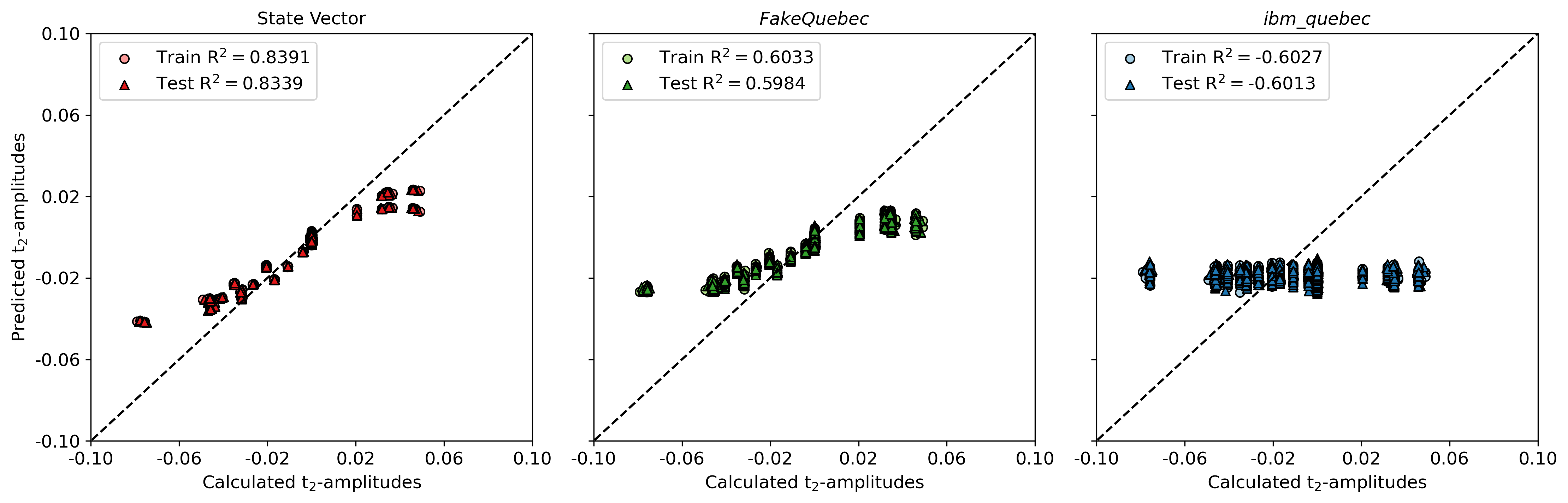}
    \caption{A comparison of the A2{\_}HWE-CNOT circuit ran using (a) state-vector simulation, (b) the noisy \textit{FakeQuebec} backend, and (c) the \textit{ibm{\_}quebec} device. The x-axis denotes the calculated, true $t_{2}$-amplitudes and the y-axis shows the predicted $t_{2}$-amplitudes.}
    \label{fig:ibmq_vs_statevector_vs_fake}
\end{figure}

Based on the model performance observed in the learning curves, we deemed the accuracy of the A2{\_}HWE-CNOT circuit sufficient for evaluation on real quantum hardware.
As mentioned in Section 2, we ported our \textsc{PennyLane} code to \textsc{Qiskit} to make use of the \textsc{Qiskit} Batch execution mode.
Before running the experiments using noisy simulation and the real hardware, we performed two forms of validation using state-vector simulation.
First, we used the optimized parameters from the A2{\_}HWE-CNOT circuit with an AL of 5 and RUD of 1, which provided the same R$^{2}$ values as in Fig.~\ref{fig:ddccRUDAL_lineplot}.
And second, since we could not use the same optimizer as we did in the \textsc{PennyLane} code, we performed an experiment using the COBYLA optimizer with 500 iterations to compare with the values in Fig.~\ref{fig:ddccRUDAL_lineplot}.
For this experiment, we reduced the training set size from 80\% to 10\% of the total available data, since training a model using 80\% of the data was computationally unfeasible based on preliminary experimentations.
As highlighted in Fig.~\ref{fig:ibmq_vs_statevector_vs_fake} (a), the method implemented in \textsc{Qiskit} offers similar, but slightly improved, performance over the \textsc{PennyLane} implementation optimized using the simultaneous perturbation stochastic approximation method (SPSA), with a training and test R$^{2}$ of $0.8391$ and $0.8339$, respectively.

Following this calibration step, we analyzed the resilience level (error mitigation), and the optimization level (related to \textsc{Qiskit}'s transpiler settings) using the \textit{FakeQuebec} backend with the \textit{qiskit-aer} plugin, as highlighted in SI Table \ref{tab:FakeQuebecOptResBenchmark}.
Based these results and on preliminary resource estimations, we determined that the maximum values we could use on the real, 127 qubit \textit{ibm{\_}quebec} device are a resilience level of 1, which uses Twirled Readout Error eXtinction (TREX) \cite{van_den_berg_model-free_2022} error mitigation, an optimization level of 3, which corresponds to the highest circuit optimization available, and 3072 shots.
The results of this model are highlighted in Fig.~\ref{fig:ibmq_vs_statevector_vs_fake} (b),
where the training set shows a decrease in accuracy, when compared to the state-vector simulations, with an R$^{2}$ of 0.6033 and a test R$^{2}$ of 0.5984.
After evaluating the resilience and optimization level using the \textit{FakeQuebec} backend, the next step is to prep the data for running on the real device.

To efficiently run the DDCC model on \textit{ibm{\_}quebec}, the data is split into batches of 4 molecules per sample, where each molecule has 64 $t_{2}$-amplitudes.
Batching the data allows for a reduced number of circuit executions per primitive unified bloc (PUB), which has a maximum of 10 million evaluations per circuit, considering the number of observables, parameters, and shots.
Using the \textit{ibm{\_}quebec} device, 500 training iterations took 24.5 days of wall clock time, which includes time waiting in the queue, device execution time, and query time to and from the IBM Quantum platform. 
Overall, each training iteration took an average of 1448.13 s on the QPU for a total of 8.38 days of QPU time.
Additionally, as highlighted in Fig.~\ref{fig:ibmq_vs_statevector_vs_fake} (c), both state-vector simulation and the \textit{FakeQuebec} backend outperform the real device, where the \textit{ibm{\_}quebec} device provided a train and test R$^{2}$ of -0.6027 and -0.6013, respectively.
This highlights the difficulty of applying the DDCC method on real quantum hardware since the query time to the IBM Quantum platform, the number of training points (proportional to $N_{\text{occ}}^{2}N_{\text{virt}}^{2}$ $t_{2}$-amplitudes), and the potential volume of device executions stemming from varying shots, device noise, resilience levels, and optimization settings must be considered.

\section{Discussion}
Overall, this study has introduced arguments against several key claims for using PQCs for regression-based tasks that must be addressed.
The first claim we evaluated was that as the number of training parameters increases, the model performance increases.
As highlighted in Figs. \ref{fig:RudAlLinePlot} and \ref{fig:RudAlParity}, for both the BSE49 and DDCC data, we found that this was not always true.
Model improvements were negligible for the 5 qubit BSE49 data, and for the DDCC data, the models plateaued after a given depth.
In the case of DDCC,  claims that data re-uploading\cite{perez-salinas_data_2020,suzuki_predicting_2020} increases the model performance do not hold since the number of ALs was found to improve model performance better than the RUD.
Furthermore, we explore the claims that increasing input redundancy\cite{gil_vidal_input_2020,suzuki_predicting_2020} increases model performance.
In the Supplementary Information, we analyze the initial encoding strategy of one qubit per feature (Fig.~\ref{fig:qiskit_statevector_0.1}) versus the same circuit using two qubits per feature (Fig.~\ref{fig:3W_qiskit_statevector_0.1}).
We found that the model consisting of two qubits per feature takes more iterations to optimize and offers poorer performance than the original A2{\_}HWE-CNOT model.
This insight is particularly informative, as all the classical analogues have training and test R$^{2}$ greater than or equal to 0.9872.
Furthermore, claims regarding the expressive power of PQCs over classical models using reduced training set sizes\cite{hatakeyama-sato_quantum_2023} are not valid for the BSE49 and DDCC datasets.
We found that the PQCs using the BSE49 dataset offer fewer overfit models, or equivalent fitness in the case of DDCC models, than their classical counterparts.
In general, both datasets are outperformed by their classical counterparts in terms of model accuracy.
Additionally, we will note that for other classes of PQCs, not examined in this study, these trends may not hold, and an ``optimal'' model may exist using different circuits.

A ``quantum enhancement'' may exist for the circuits we examined herein regarding the time complexity required to train the quantum circuit classically versus on the quantum hardware. 
It is important to note that certain quantum circuits---such as Instantaneous Quantum Polynomial-time (IQP) circuits---are often cited as offering a form of “quantum advantage” due to their theoretical intractability for classical simulation. This advantage, however, refers specifically to computational hardness in a worst-case sense and does not inherently imply that these circuits will outperform classical machine learning models in practical applications or on real-world datasets.\cite{lund_quantum_2017,harrow_quantum_2017}
This highlights the importance of formulating the machine learning task to take advantage of the inherent quantum nature of quantum computing.
This could include creating better molecular representations specifically tailored to QML-related tasks.
This challenge was highlighted with the BSE49 data, which performed poorly using both 5 and 16 qubits. 
One of the drawbacks of using molecular representations that encode structural information is that a high level of compression is required, i.e., reducing a vector of 2048 features per molecule down to 5/16 features representing the principle components of the original feature vector.
Another possibility, discussed in \citet{suzuki_predicting_2020}, is the integration of PQCs with classical neural networks, where the PQC creates a quantum-enhanced feature mapping used as input for the classical model.
Lastly, one potential avenue for improving the DDCC model is to reformulate the problem by incorporating additional features that capture more aspects of quantum simulation. This could help shift the problem further from its classical analogue, potentially making it more suitable for quantum machine learning approaches.
Additionally, models that exclude small $t_{2}$-amplitudes, such as those found in \citet{pathirage_exploration_2024}, could be used to increase performance but are out of the scope of this work.

\section{Conclusions}
To summarize, this work explores the applicability of parametrized quantum circuits (PQCs) to problems in computational chemistry by evaluating their performance on two chemically relevant datasets: the BSE49 dataset of bond separation energies and a set of water conformers generated using the data-driven coupled-cluster (DDCC) method. 
We began with a comprehensive assessment of 168 PQCs, constructed from various combinations of data encoding and ansatz layers, and identified the most promising 5-qubit circuit for each dataset. 
From a practical perspective, none of the methods modeled the BSE49 dataset in a useful way, since R$^{2}$ values of close to zero or even negative R$^{2}$ values (poorer performance than using the mean) were obtained. 
This poor performance might stem from the information loss that resulted from condensing the molecular representation into a very small number of qubits. 
In contrast, PQCs generated actual predictions for the $t_{2}$-amplitudes of the water conformers problem using the DDCC method, with R$^{2}$ for the best method of approximately 0.60.  
We then investigated how increasing circuit depth and varying the size of the training set impact predictive accuracy, which increased the R$^{2}$ for the water conformers/DDCC problem to \textit{ca.} 0.8, arguably within the potentially useful range. 
This good performance is likely due to negligible information loss (compression) when encoding the problem onto qubits, but is also consistent with the idea that intrinsically quantum properties might benefit from a quantum method used in their simulation. 
The best-performing PQC was further evaluated under realistic conditions, using both noisy simulations and actual quantum hardware. 
The usefulness of the $t_{2}$-amplitude prediction went away (R$^{2}$ drastically reduced) when real quantum hardware was used, indicating that noise levels on real quantum hardware are still too high. 
Lastly, we reflected on the current limitations of PQCs in achieving practical quantum advantage for chemically motivated problems and proposed potential strategies to enhance their relevance and performance in future quantum machine learning studies.


\section{Acknowledgements}
We acknowledge the Government of Canada’s New Frontiers in Research Fund (NFRF), for grant NFRFE-2022-00226, and the Quantum Software Consortium (QSC), financed under grant \#ALLRP587590-23 from the National Sciences and Engineering Research Council of Canada (NSERC) Alliance Consortia Quantum Grants.
This research was enabled in part by computational support provided by IBM Quantum via the Quantum Software Consortium and PINQ2, along with access to classical resources through the Digital Research Alliance of Canada.
The authors also acknowledge Aviraj Newatia, who provided valuable insights based on preliminary results that are not included in this work, and Nick Taylor, who supported the initial stages of the project and received support from the Centre for Quantum Information and Quantum Control (CQIQC) at the University of Toronto.
Additionally, Viki Kumar Prasad would like to acknowledge the Data Sciences Institute (DSI) for its support via the DSI Postdoctoral Fellowship throughout this project.

\section{Data and Software Availability}
All code and data used in this study are hosted on GitHub, free of charge at \href{https://github.com/MSRG/qregress/}{https://github.com/MSRG/qregress/}.

\section{Supplementary Information}
Information regarding the feature reduction, regression heatmaps, parity plots of the ansatz layers (ALs) and re-upload depth (RUD), learning curve data, \textit{FakeQuebec} results, and information regarding the input data redundancy can be found in the Supplementary Information.

\bibliography{achemso-demo}

\end{document}


\newpage
\section*{Electronic Supplementary Information}
\setcounter{page}{1}
\renewcommand{\thepage}{S-\arabic{page}}

\subsection*{Table of Contents}
\begin{table}[H]
	\centering
	\begin{tabular}{|c|c|c|}
		\hline
		\ref{section:BSE49_Feature_Set} & Classical Feature Reduction& \pageref{section:BSE49_Feature_Set} \\ 
		\hline
            \ref{section:DDCC_Feature_Set} & DDCC Feature Set & \pageref{section:DDCC_Feature_Set} \\
            \hline
            \ref{section:ALRUDExp} & Ansatz Layer and Re-Upload Depth Experiments & \pageref{section:ALRUDExp} \\
            \hline
            \ref{section:5BSE49_LC_data} & BSE49 5 Qubit Learning Curve Data & \pageref{section:5BSE49_LC_data} \\
            \hline
            \ref{section:5DDCC_LC_data} & DDCC 5 Qubit Learning Curve Data & \pageref{section:5DDCC_LC_data} \\
            \hline            
		\ref{section:DDCC_fake}  & DDCC Fake Quebec & \pageref{section:DDCC_fake} \\
		\hline
		\ref{section:ddcc_data_redundancy}  & DDCC Data Redundancy & \pageref{section:ddcc_data_redundancy} \\
		\hline        
	\end{tabular}
	\label{tab:my_label}
\end{table}

\setcounter{table}{0}
\renewcommand{\tablename}{Table}
\renewcommand{\thetable}{S\arabic{table}}

\setcounter{figure}{0}
\renewcommand{\figurename}{Figure}
\renewcommand{\thefigure}{S\arabic{figure}}

\setcounter{section}{0}
\renewcommand{\thesection}{S\arabic{section}}
\newpage

\section{BSE49 Feature Set}\label{section:BSE49_Feature_Set}
\begin{figure}[H]
	\centering	
	\includegraphics[width=\linewidth]{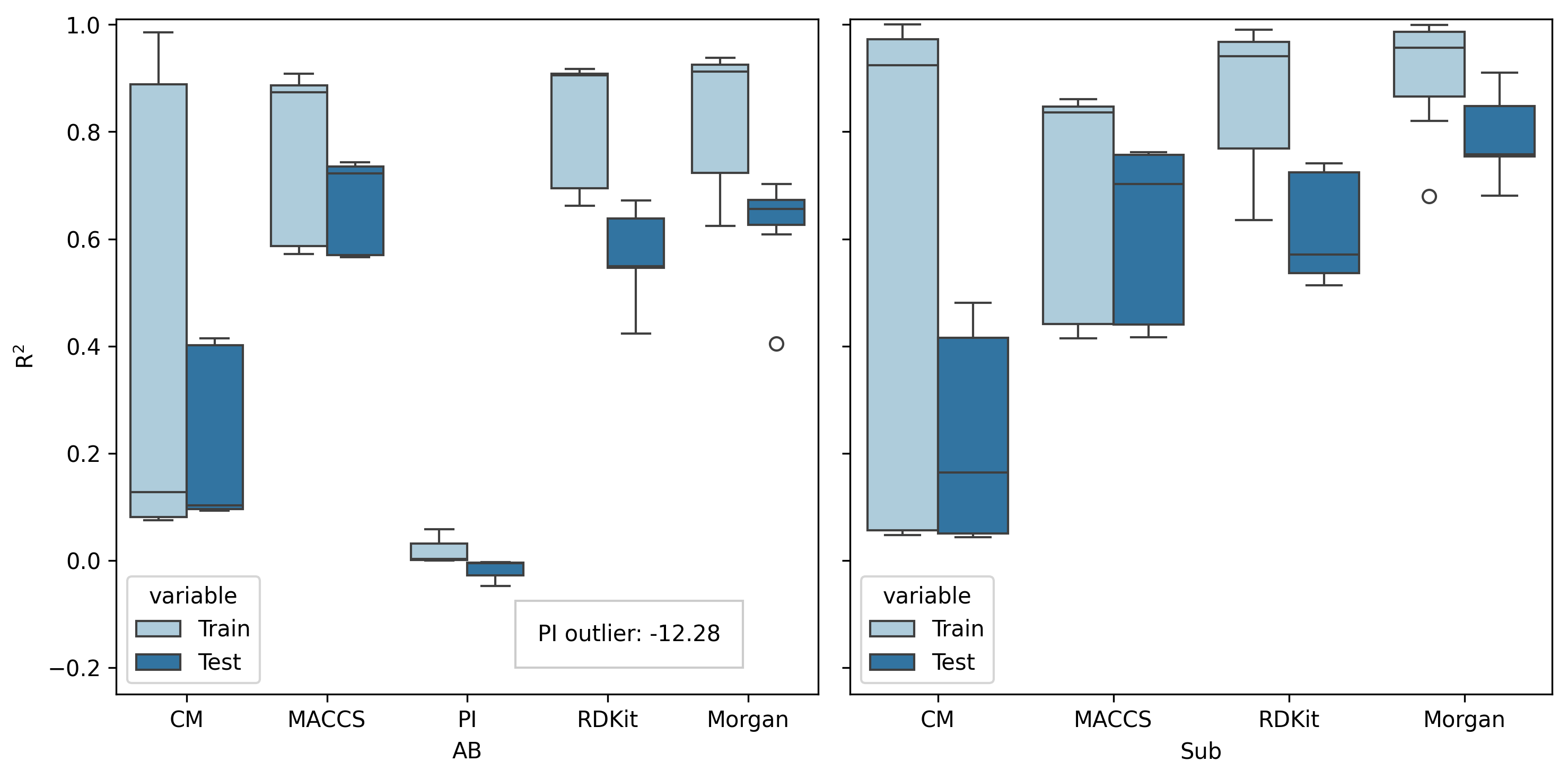}
	\caption{Coulomb matrices (CMs), Molecular ACCess Systems (MACCS), persistence images (PIs), RDKit and Morgan fingerprints. Performance of a diverse set of molecular representations R$^{2}$}
	\label{fig:classical_molrepfig}
\end{figure}

Using a diverse set of molecular representations discussed in the main text, we examined various classical regression models to help choose the optimal representation for our diverse set of PQCs.
The classical models we analyze include ridge, lasso, elastic net, \textit{k}-nearest-neighbors, random forest, gradient boosting, support vector machines, kernel ridge, and Gaussian process regression as implemented in sckit-learn.\cite{pedregosa_scikit-learn_2011}
As highlighted in Fig. \ref{fig:classical_molrepfig}, we found that the best molecular representation across all models tested was Morgan fingerprints using the \textit{sub} formulation, which we use for all experiments using the PQCs.

\begin{figure}[H]
	\centering	
	\begin{subfigure}[b]{0.49\textwidth}
		\centering
		\includegraphics[width=\textwidth]{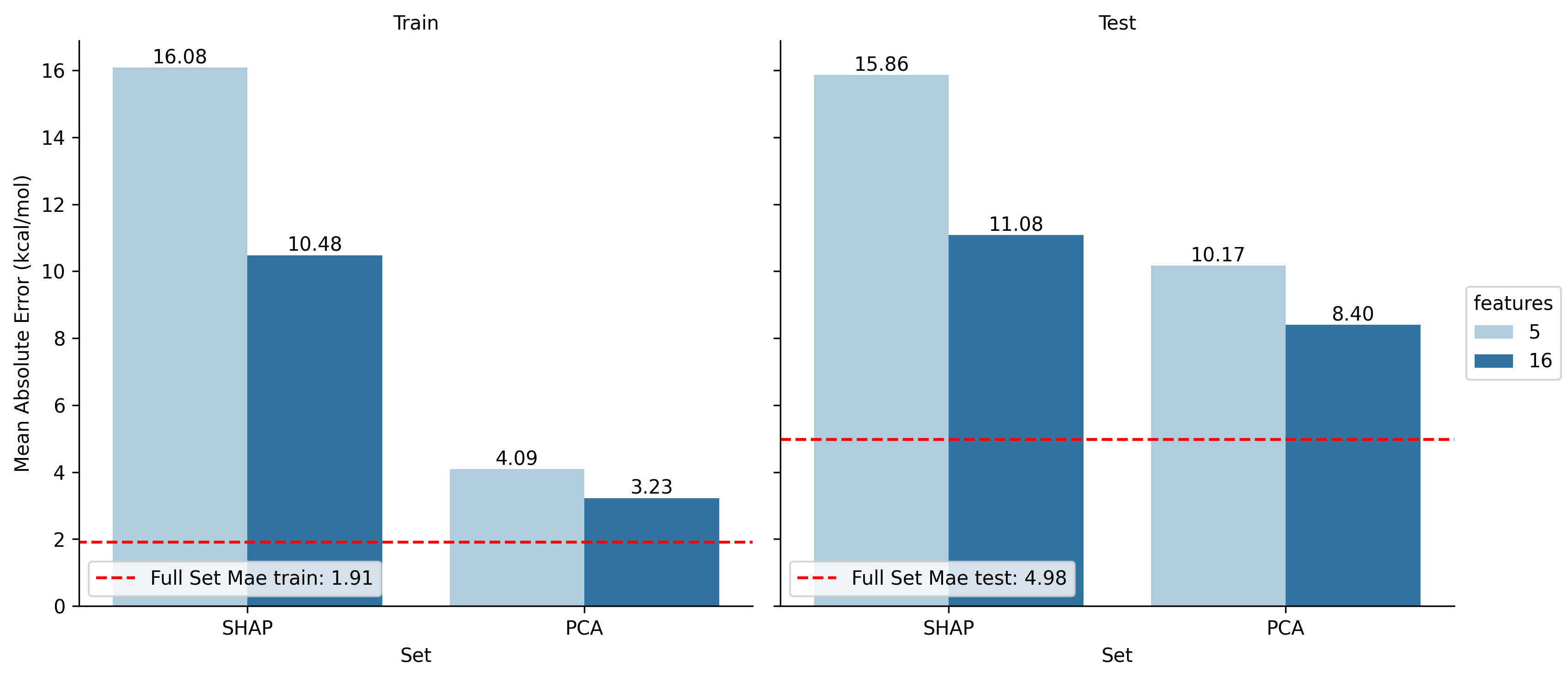}
		\caption{}
		\label{fig:BSE_bse_classical_features_MAE}
	\end{subfigure}
	\hfill		
	\begin{subfigure}[b]{0.49\textwidth}
		\centering
		\includegraphics[width=\textwidth]{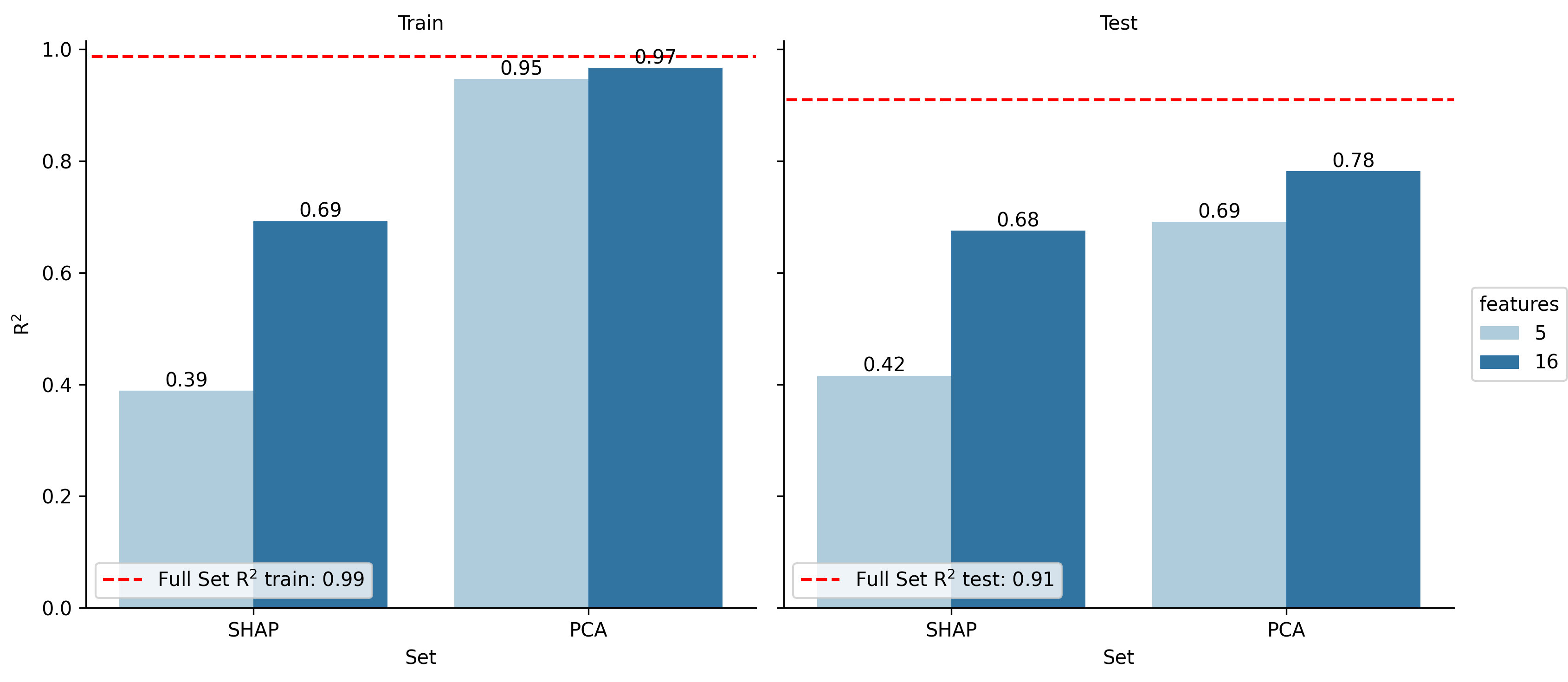}
		\caption{}
		\label{fig:BSE_classical_features_R2}
	\end{subfigure}
	\caption{Feature reduction of the BSE dataset represented using SHapley Additive ExPlanation analysis (SHAP) and principal component analysis (PCA) for the training set (left) and test set (right) using both 5 (light blue) and 16 (dark blue) features evaluated using the (a) mean absolute error (in kcal/mol) and (b) R$^{2}$. The dashed red line is used to denote the classical baseline model.}
	\label{fig:bse_classical_features}
\end{figure}

Additionally, one challenge of applying classical molecular representations for ML tasks using PQCs is reducing the set of features, often containing hundreds or thousands of features per sample, to the number of qubits used on the quantum device.
Due to the cost of state-vector simulations as circuit depth increases, we reduce the initial set of features from Morgan fingerprints from 2048 features down to 5 or 16 features.
To perform this reduction, we explored two different methods, SHapley Additive ExPlanation analysis (SHAP)\cite{lundberg_unified_2017}, which uses cooperative game theory to determine feature importance, and principal component analysis (PCA), which is a statistical technique for dimensionality reduction that works by identifying the directions (principal components) along which the variance in the dataset is maximized, as implemented in scikit-learn.\cite{pedregosa_scikit-learn_2011}
In Figs. \ref{fig:BSE_classical_features_R2} and \ref{fig:BSE_bse_classical_features_MAE}, we compare the two reduction techniques for both 5 and 16 features, with the initial model containing 2048 features.
The initial model has mean absolute errors (MAE) of 1.91 and 4.98 kcal/mol and R$^{2}$s of 0.99 and 0.91 for the training and test set, respectively.
Using SHAP, we found that the model containing 5 features has an MAE of 16.08 kcal/mol and an R$^{2}$ of 0.39 for the training set, while the test set has an MAE of 15.86 kcal/mol and an R$^{2}$ of 0.42.
For the model with 16 features, we observe improvements in both the training and test sets, with MAEs of 10.48 and 11.08 kcal/mol and R$^{2}$ values of 0.69 and 0.68, respectively.
For the models using 5 and 16 features reduced through PCA, the training sets yield MAEs of 4.09 and 3.23 kcal/mol, while the test sets have MAEs of 10.17 and 8.40 kcal/mol, respectively.
The R$^{2}$s also show slight improvement over the feature reductions using SHAP, where the R$^{2}$s using 5 features are 0.95 and 0.69 for the training and test set, respectively, and the model using 16 features has R$^{2}$s of 0.97 and 0.78 for the training and test set, respectively.
Overall, the models that use SHAP for feature reduction offer less overfit models while sacrificing accuracy when compared to the models reduced using PCA, which offer more accurate models that are more overfit.

\section{DDCC Feature Set}\label{section:DDCC_Feature_Set}
Herein, we provide a brief overview of the equations required to understand the DDCC method, starting with the coupled-cluster wave function, which takes the general form,
\begin{equation}
	\ket{\Psi_{\text{CC}}} = \exp(\hat{T}) \ket{\Psi_{0}}
	\label{eq:cc_wfn}
\end{equation}
where $\hat{T}$ is the cluster operator and $\ket{\Psi_{0}}$ is the reference (usually Hartree-Fock) wave function.
In CCSD, the $\hat{T}$ operator is truncated to only include single ($\hat{T}_{1}$) and double ($\hat{T}_{2}$) excitations.
After solving the coupled-cluster equations, the CCSD correlation energy can be formulated as,
\begin{equation}
	E^{\text{CCSD}}_{\text{corr}} = \sum_{\substack{a<b \\ i<j}} \mel{ij}{}{ab} t^{ab}_{ij} + \sum_{\substack{a<b \\ i<j}} \mel{ij}{}{ab} t^{a}_{i} t^{b}_{j}
	\label{eq:cc_corr}
\end{equation}
where $i$ and $j$ denote occupied orbitals, $a$ and $b$ denote virtual orbitals, $t^{ab}_{ij}$ correspond to two-electron excitation amplitudes ($t_{2}$-amplitudes), $t^{a}_{i}$ and $t^{b}_{j}$ correspond to one-electron excitation amplitudes ($t_{1}$-amplitudes), and $\mel{ij}{}{ab}$ are two-electron integrals.

The objective of the DDCC method is to learn the CCSD $t_{2}$-amplitudes using features generated using HF and MP2, since the CCSD $t_{2}$-amplitudes are initialized using MP2 $t_{2}$-amplitudes, defined as,
\begin{equation}
	t^{ab}_{ij(\text{MP2})} = \frac{\mel{ij}{}{ab}}{\varepsilon_{i}+\varepsilon_{j}-\varepsilon_{a}-\varepsilon_{b}}
	\label{eq:MP2_t2}
\end{equation}
where $\varepsilon_{i}$ and $\varepsilon_{j}$ denote the orbital energies of the occupied orbitals $i$ and $j$, while the virtual orbitals $a$ and $b$ are denoted by $\varepsilon_{a}$ and $\varepsilon_{b}$.

\begin{figure}[H]
	\centering
	\includegraphics[width=0.49\textwidth]{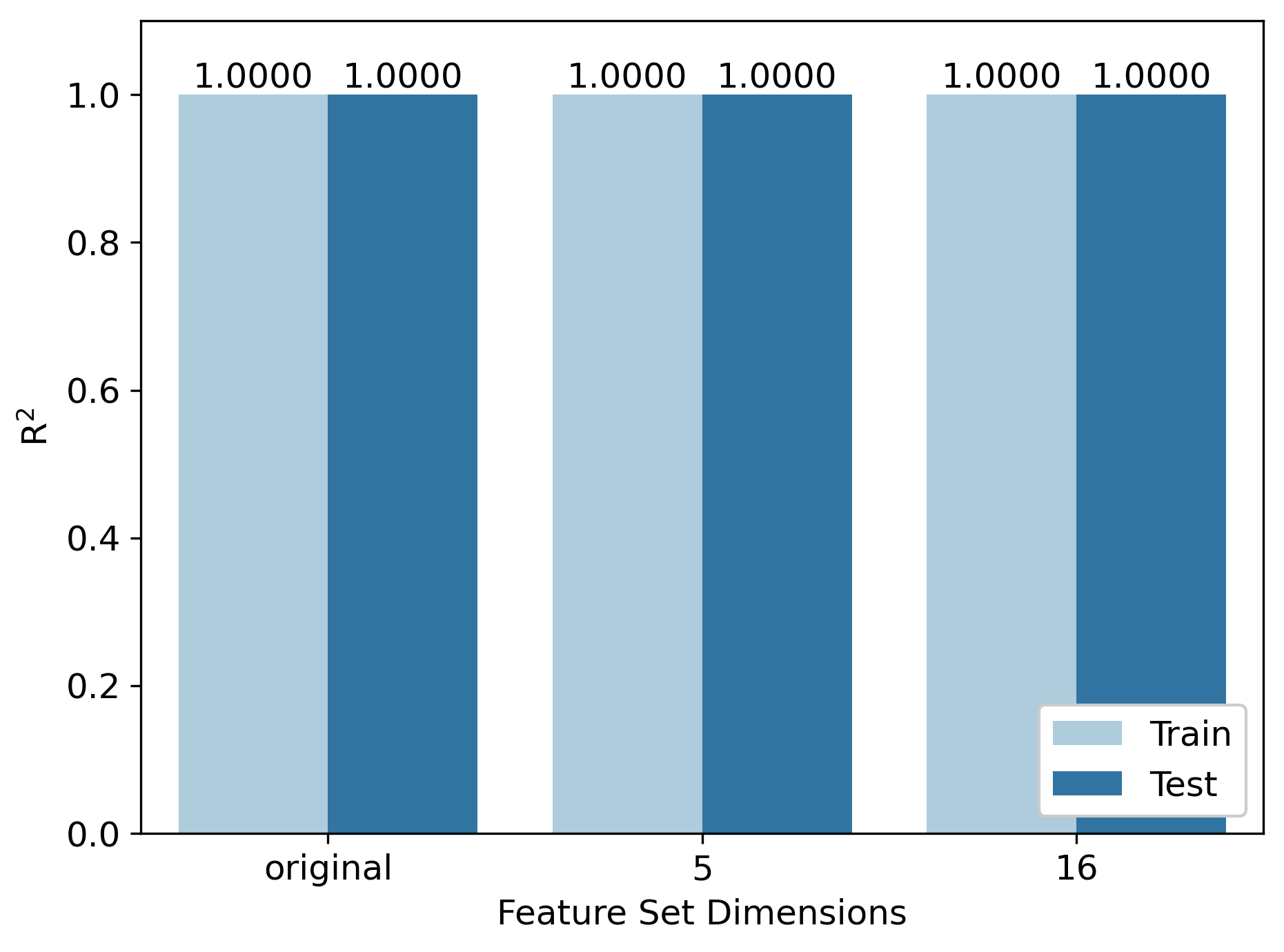}
	\caption{Feature reduction performed using SHapley Additive ExPlanation analysis (SHAP) values for the training set (light blue) and test set (dark blue) using 30 (original), 5, and 16 features.}
	\label{fig:DDCC_feature_set}
\end{figure}

Like the BSE49 dataset, the 30 DDCC features must be reduced to 5 or 16 features using either SHAP or PCA.
Unlike the BSE49 dataset, SHAP outperforms PCA due to the direct relationship between the feature set and target values.
Additionally, as highlighted in Fig. \ref{fig:DDCC_feature_set}, when the feature set is reduced to 5 or 16 features, both models have train and test R$^{2}$s of 1.00.

\section{Regression Heatmaps}
\begin{figure}[H]
	\centering	
	\begin{subfigure}[b]{0.6\textwidth}
		\centering
		\includegraphics[width=\linewidth]{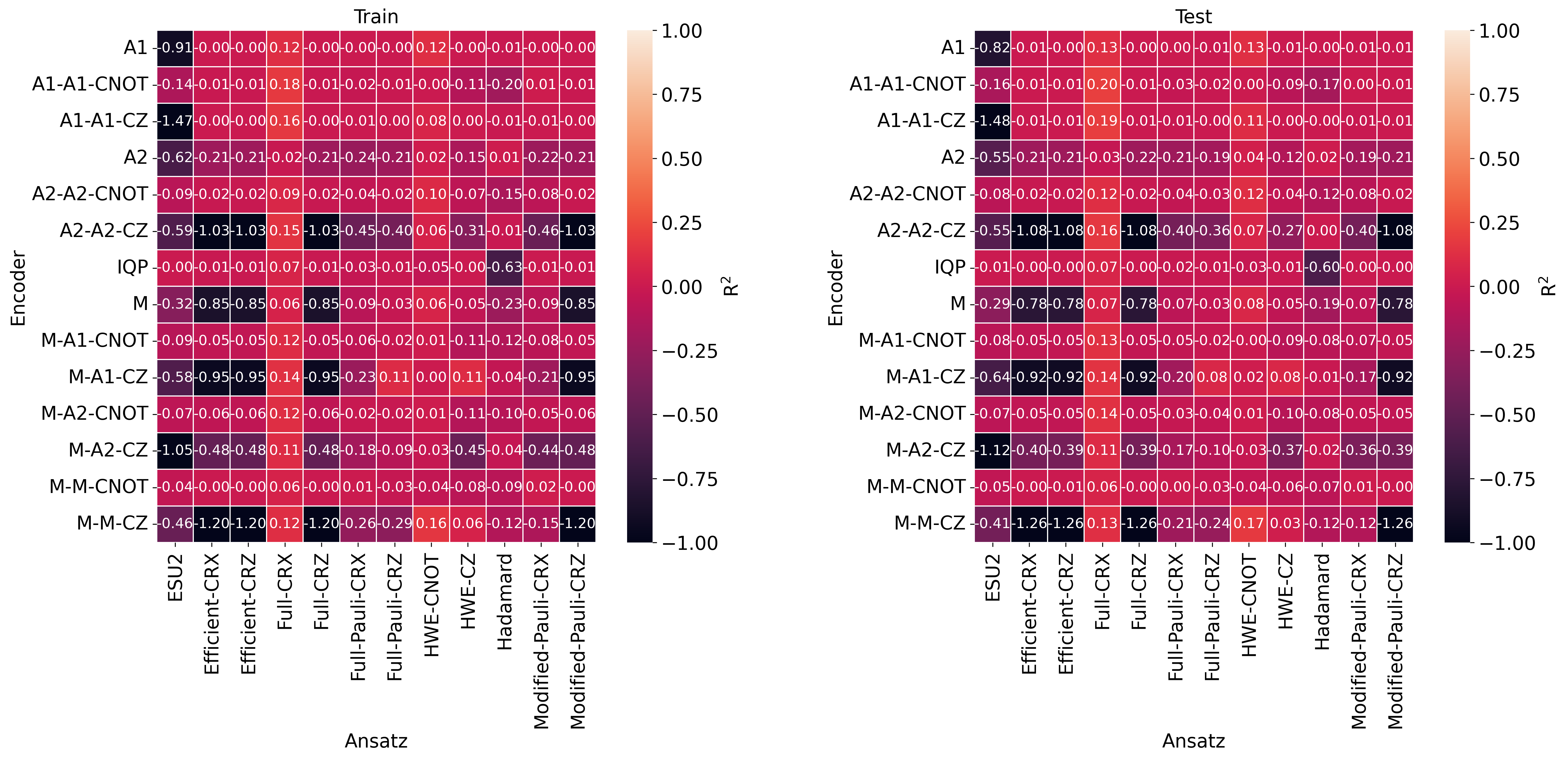}
		\caption{}
		\label{fig:5BSE_heatplots}
	\end{subfigure}
	\hfill
	\begin{subfigure}[b]{0.6\textwidth}
		\centering
		\includegraphics[width=\linewidth]{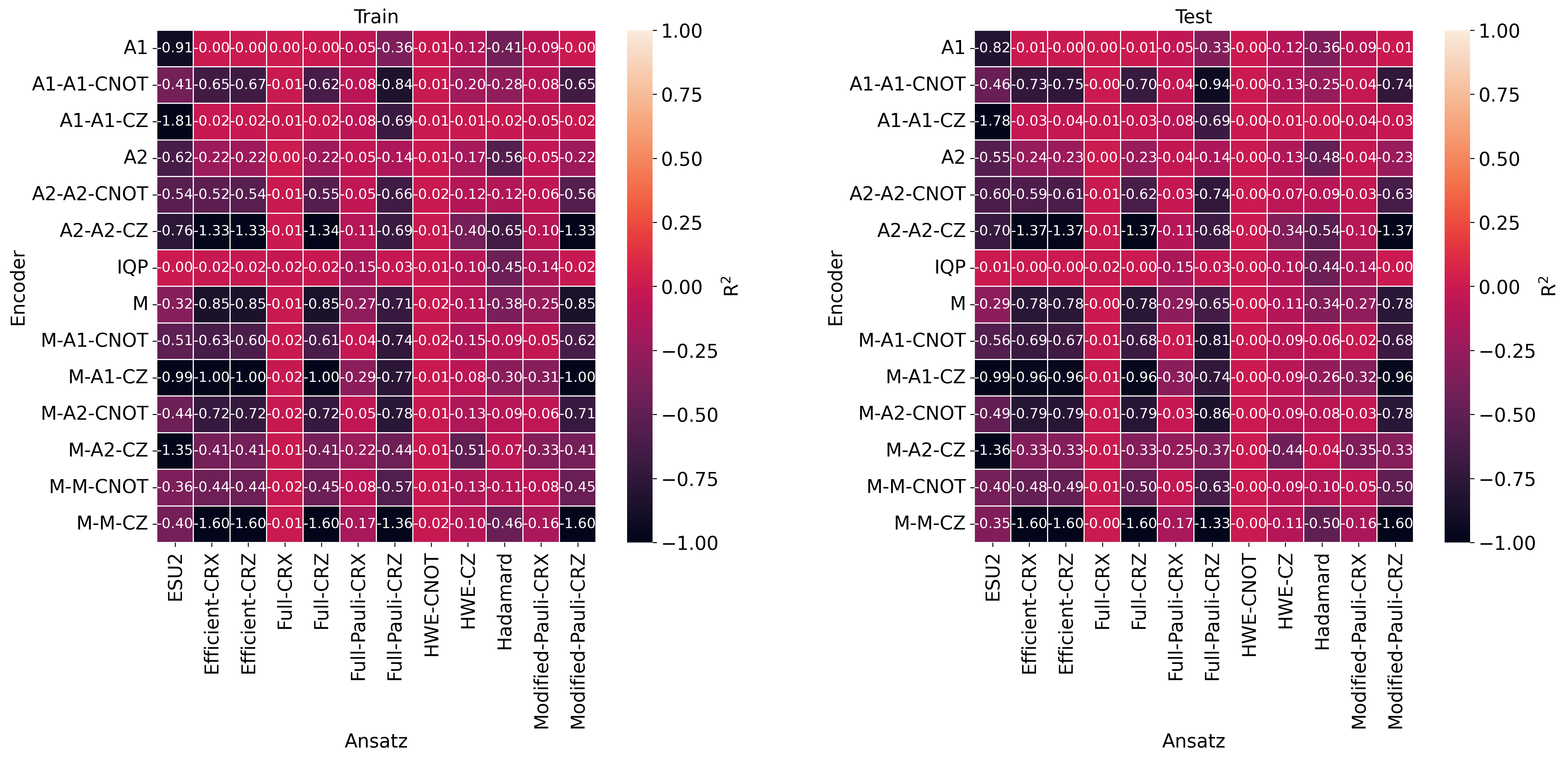}
		\caption{}
		\label{fig:16BSE_heatplots}
	\end{subfigure}
	\caption{The model performance (in R$^{2}$) of the 168 parametrized quantum circuits for BSE49 using (a) 5 and (b) 16 qubits, where the x-axis denotes the ansatz and the y-axis denotes the encoders. }
	\label{fig:BSEboxandheat}	
\end{figure}

\begin{figure}[H]
    \centering	
    \includegraphics[width=\linewidth]{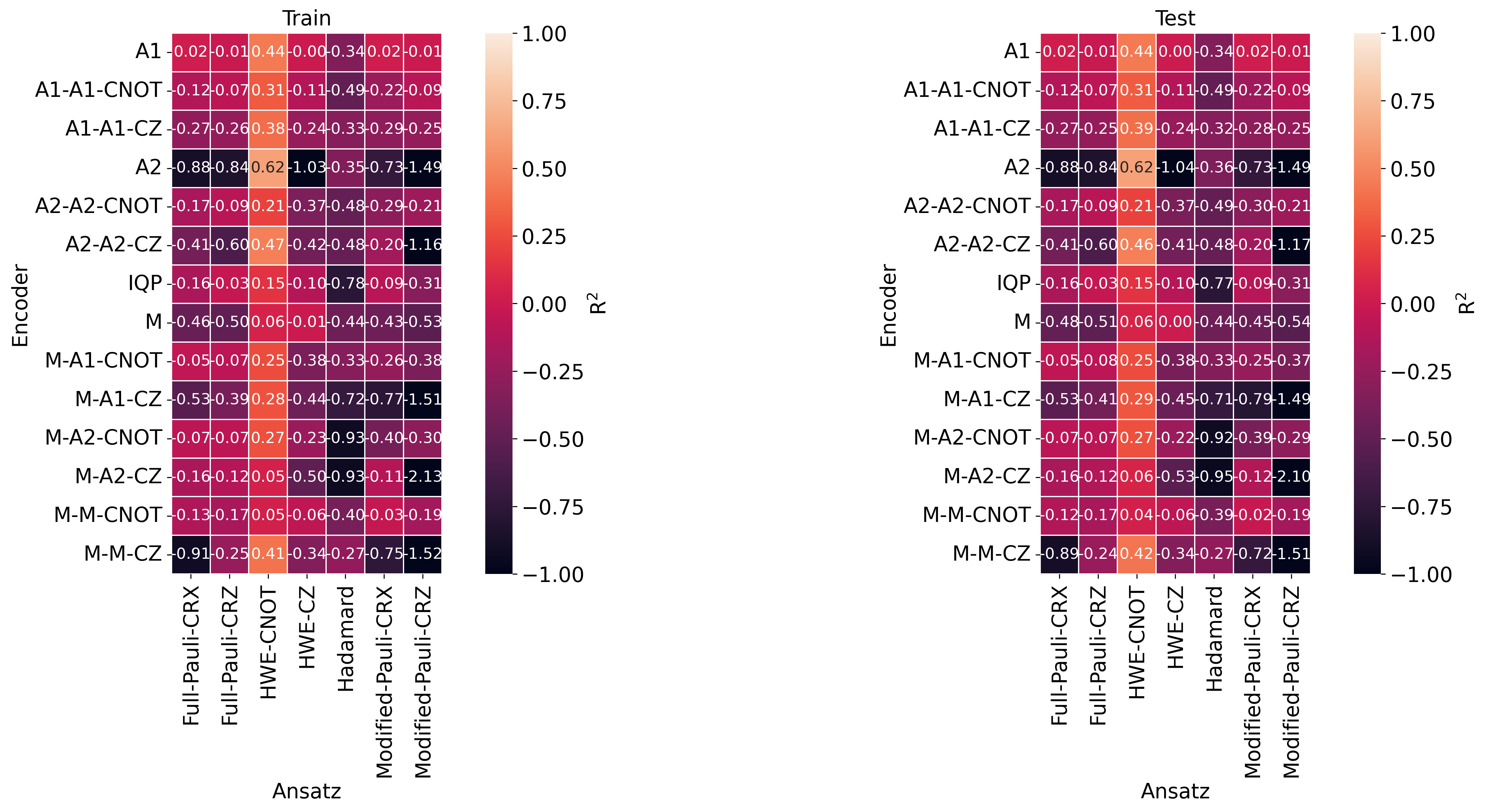}
    
    \caption{The model performance, using the coefficient of determination (R$^{2}$), for the reduced set of 98 PQCs using the DDCC dataset, using heat maps for the training (left) and test sets (right)}
    \label{fig:ddccheatplots}	
\end{figure}

\section{Ansatz Layer and Re-Upload Depth Experiments}\label{section:ALRUDExp}
\begin{figure}[H]
	\centering	
	\begin{subfigure}[b]{\textwidth}
		\centering
		\includegraphics[width=0.8\linewidth]{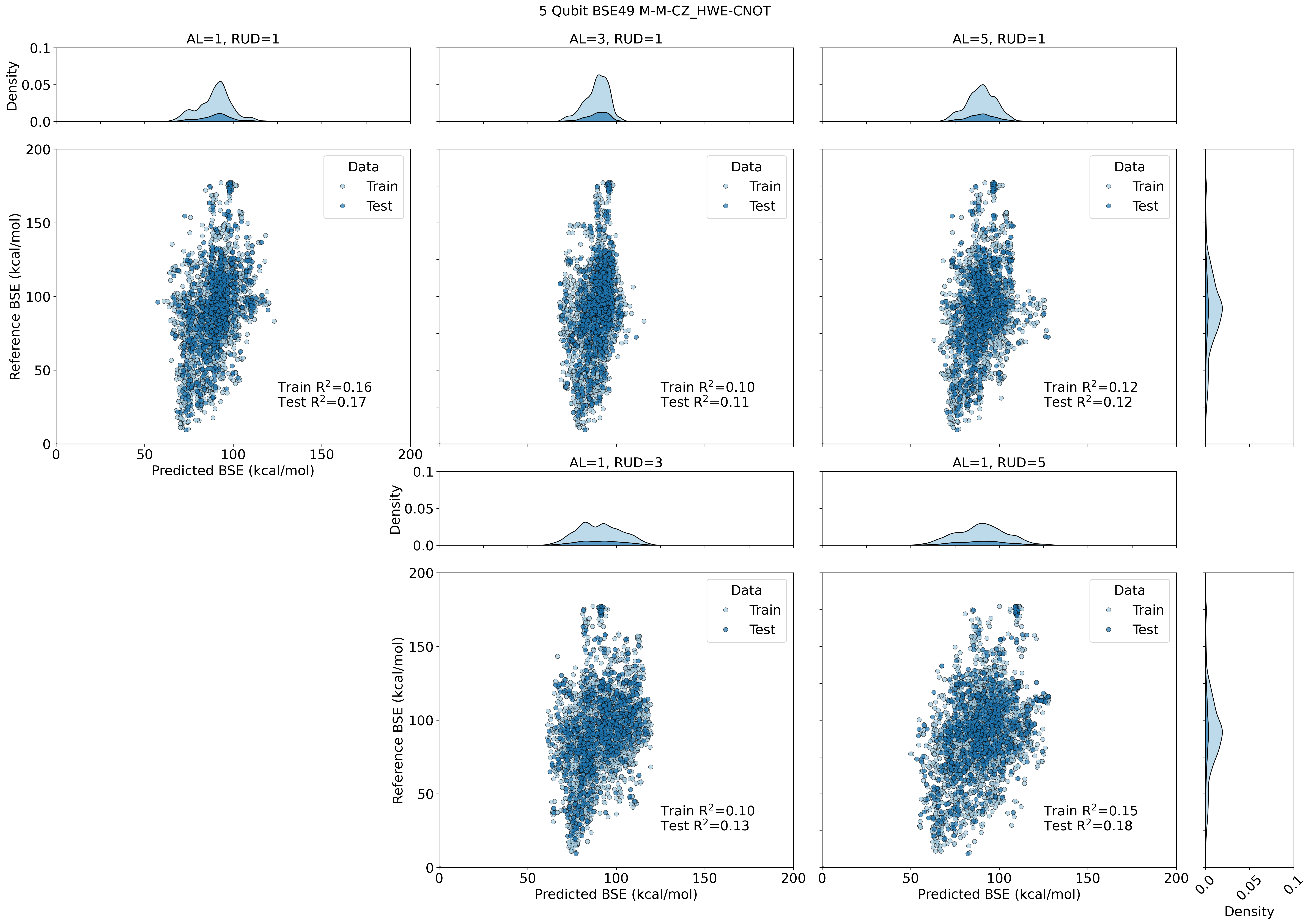}
		\caption{}
		\label{fig:BSE5_distribution_parity}
	\end{subfigure}
	\hfill

	\begin{subfigure}[b]{0.8\textwidth}
		\centering
		\includegraphics[width=\linewidth]{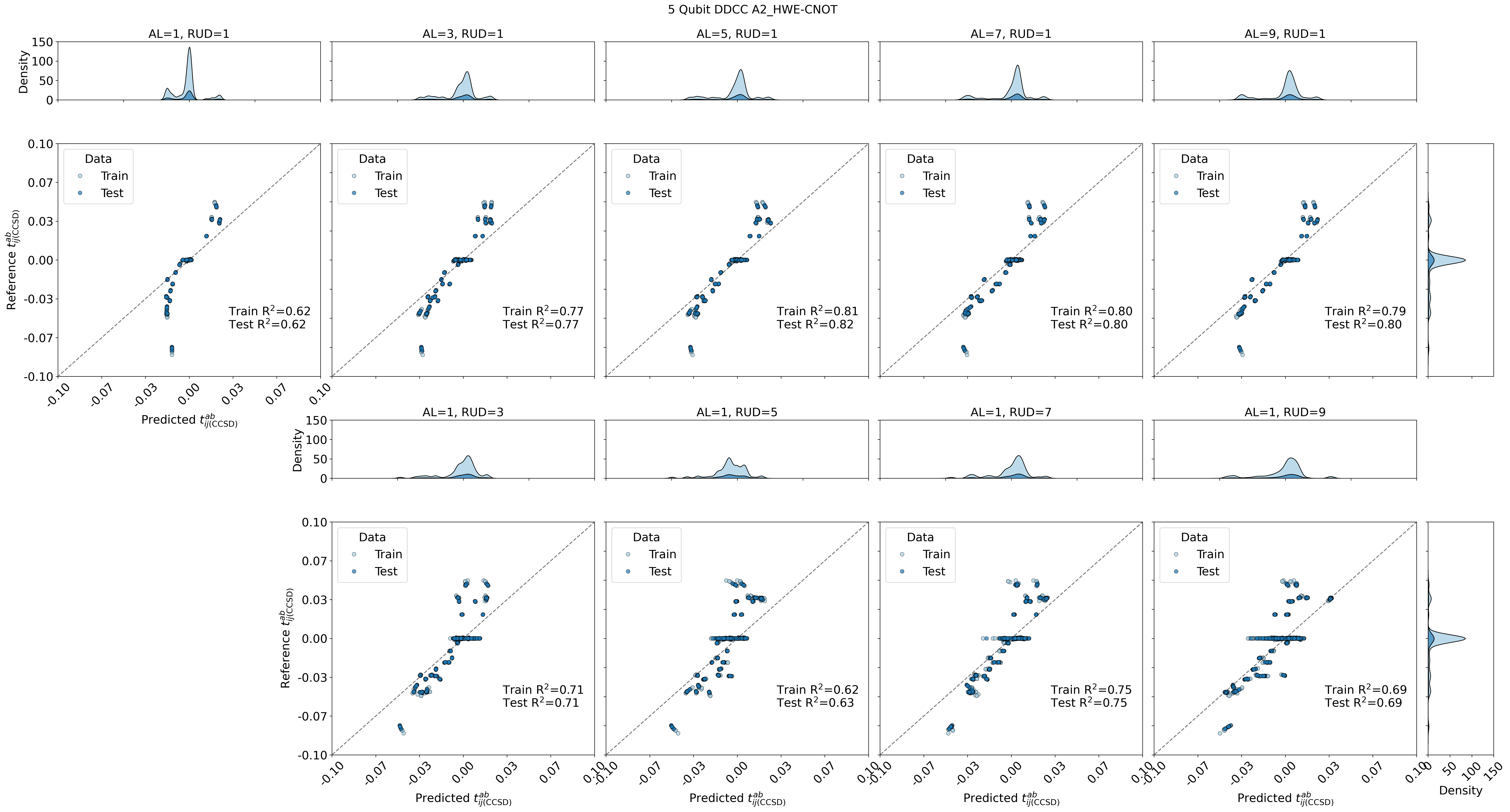}
		\caption{}
		\label{fig:ddccdistribution_parity}
	\end{subfigure}	
	\caption{Regression parity plots for each AL and RUD are shown for the (a) BSE49 and (b) DDCC data, where the x-axis denotes the predicted target values and the y-axis denotes the reference data.}	
    \label{fig:RudAlParity}
\end{figure}

\section{BSE49 5 Qubit Learning Curve Data}\label{section:5BSE49_LC_data}
\begin{longtable}{llrr}
\toprule
ratio & model &  Train & Test \\
\midrule
\multirow[t]{10}{*}{0.1} & M-M-CZ{\_}HWE-CNOT & 0.2015 & 0.1303 \\
& elastic & 0.1787 & 0.1229 \\
& gpr & 0.3254 & 0.2147 \\
& grad & 0.9010 & 0.3352 \\
& knn & 0.9998 & 0.3963 \\
& krr & 0.4242 & 0.2811 \\
& lasso & 0.1786 & 0.1229 \\
& rfr & 0.8896 & 0.4220 \\
& ridge & 0.1789 & 0.1226 \\
& svr & 0.4020 & 0.2719 \\
\cline{1-4}
\multirow[t]{10}{*}{0.3} & M-M-CZ{\_}HWE-CNOT & -0.0172 & -0.0088 \\
& elastic & 0.2099 & 0.2012 \\
& gpr & 0.4211 & 0.3857 \\
& grad & 0.8896 & 0.5215 \\
& knn & 0.9998 & 0.5837 \\
& krr & 0.4967 & 0.4385 \\
& lasso & 0.2094 & 0.2002 \\
& rfr & 0.9455 & 0.5589 \\
& ridge & 0.2100 & 0.2015 \\
& svr & 0.4881 & 0.4287 \\
\cline{1-4}
\multirow[t]{10}{*}{0.5} & M-M-CZ{\_}HWE-CNOT & 0.1715 & 0.1792 \\
& elastic & 0.1960 & 0.2050 \\
& gpr & 0.4383 & 0.4377 \\
& grad & 0.9350 & 0.5721 \\
& knn & 0.9998 & 0.6222 \\
& krr & 0.4724 & 0.4612 \\
& lasso & 0.1954 & 0.2039 \\
& rfr & 0.9453 & 0.5970 \\
& ridge & 0.1961 & 0.2055 \\
& svr & 0.4754 & 0.4628 \\
\cline{1-4}
\multirow[t]{10}{*}{0.7} & M-M-CZ{\_}HWE-CNOT & 0.1908 & 0.2064 \\
& elastic & 0.1623 & 0.1785 \\
& gpr & 0.4347 & 0.4516 \\
& grad & 0.9624 & 0.5957 \\
& knn & 0.9998 & 0.6366 \\
& krr & 0.4577 & 0.4714 \\
& lasso & 0.1617 & 0.1770 \\
& rfr & 0.9483 & 0.6430 \\
& ridge & 0.1623 & 0.1791 \\
& svr & 0.4577 & 0.4709 \\
\cline{1-4}
\multirow[t]{10}{*}{0.8} & M-M-CZ{\_}HWE-CNOT & 0.1559 & 0.1727 \\
& elastic & 0.1656 & 0.1828 \\
& gpr & 0.4445 & 0.4607 \\
& grad & 0.9531 & 0.6030 \\
& knn & 0.9999 & 0.6725 \\
& krr & 0.4596 & 0.4727 \\
& lasso & 0.1651 & 0.1812 \\
& rfr & 0.9462 & 0.6400 \\
& ridge & 0.1657 & 0.1834 \\
& svr & 0.4578 & 0.4671 \\
\cline{1-4}
\bottomrule
\label{table:5BSE49_LC}
\end{longtable}

\section{DDCC 5 Qubit Learning Curve Data}\label{section:5DDCC_LC_data}

\begin{longtable}{llrr}
	\toprule
	ratio & model & Test & Train \\
	\midrule
	\midrule
        \multirow[t]{10}{*}{0.100000} & A2{\_}HWE-CNOT & 0.7921 & 0.7824 \\
         & elastic & 0.9874 & 0.9877 \\
         & gpr & 1.0000 & 0.9997 \\
         & grad & 1.0000 & 0.9999 \\
         & knn & 1.0000 & 1.0000 \\
         & krr & 0.9999 & 0.9997 \\
         & lasso & 0.9873 & 0.9877 \\
         & rfr & 1.0000 & 0.9999 \\
         & ridge & 0.9898 & 0.9901 \\
         & svr & 0.9985 & 0.9982 \\
        \cline{1-4}
        \multirow[t]{10}{*}{0.300000} & A2{\_}HWE-CNOT & 0.7711 & 0.7643 \\
         & elastic & 0.9873 & 0.9875 \\
         & gpr & 1.0000 & 1.0000 \\
         & grad & 1.0000 & 1.0000 \\
         & knn & 1.0000 & 1.0000 \\
         & krr & 0.9999 & 0.9999 \\
         & lasso & 0.9872 & 0.9874 \\
         & rfr & 1.0000 & 1.0000 \\
         & ridge & 0.9897 & 0.9899 \\
         & svr & 0.9990 & 0.9989 \\
        \cline{1-4}
        \multirow[t]{10}{*}{0.500000} & A2{\_}HWE-CNOT & 0.7720 & 0.7718 \\
         & elastic & 0.9875 & 0.9875 \\
         & gpr & 1.0000 & 1.0000 \\
         & grad & 1.0000 & 1.0000 \\
         & knn & 1.0000 & 1.0000 \\
         & krr & 0.9999 & 1.0000 \\
         & lasso & 0.9875 & 0.9874 \\
         & rfr & 1.0000 & 1.0000 \\
         & ridge & 0.9899 & 0.9899 \\
         & svr & 0.9994 & 0.9994 \\
        \cline{1-4}
        \multirow[t]{10}{*}{0.700000} & A2{\_}HWE-CNOT & 0.7715 & 0.7660 \\
         & elastic & 0.9874 & 0.9875 \\
         & gpr & 1.0000 & 1.0000 \\
         & grad & 1.0000 & 1.0000 \\
         & knn & 1.0000 & 1.0000 \\
         & krr & 1.0000 & 0.9999 \\
         & lasso & 0.9874 & 0.9875 \\
         & rfr & 1.0000 & 1.0000 \\
         & ridge & 0.9898 & 0.9899 \\
         & svr & 0.9995 & 0.9995 \\
        \cline{1-4}
        \multirow[t]{10}{*}{0.800000} & A2{\_}HWE-CNOT & 0.7972 & 0.7997 \\
         & elastic & 0.9875 & 0.9874 \\
         & gpr & 1.0000 & 1.0000 \\
         & grad & 1.0000 & 1.0000 \\
         & knn & 1.0000 & 1.0000 \\
         & krr & 0.9999 & 1.0000 \\
         & lasso & 0.9874 & 0.9873 \\
         & rfr & 1.0000 & 1.0000 \\
         & ridge & 0.9899 & 0.9898 \\
         & svr & 0.9995 & 0.9995 \\
        \cline{1-4}
        \bottomrule
        \label{table:DDCC_LC}
\end{longtable}

\section{DDCC Fake Quebec}\label{section:DDCC_fake}
\setcounter{table}{0}
\begin{table}[H]
    \centering
    \begin{tabular}{|c|c|c|c|}
    \toprule
    Opt. Level & Res. Level & R$^{2}$ Train & R$^{2}$ Test \\
    \midrule
    0 & 0 & -0.0806 & -0.0882 \\
    1 & 0 & 0.5527 & 0.5447 \\
    2 & 0 & 0.5659 & 0.5578 \\
    3 & 0 & 0.5248 & 0.5186 \\
    0 & 1 & -0.0758 & -0.0874 \\
    1 & 1 & 0.5018 & 0.4939 \\
    2 & 1 & 0.5597 & 0.5565 \\
    3 & 1 & 0.5830 & 0.5750 \\
    0 & 2 & 0.1798 & 0.1824 \\
    1 & 2 & 0.4570 & 0.4527 \\
    2 & 2 & 0.5237 & 0.5207 \\
    3 & 2 & 0.5817 & 0.5782 \\
    \bottomrule
    \end{tabular}

    \caption{Analysis of the different optimization and resilience (error mitigation) levels regarding the R$^{2}$s of the training and test sets.}
    \label{tab:FakeQuebecOptResBenchmark}
\end{table}

\section{DDCC Data Redundancy}\label{section:ddcc_data_redundancy}
\begin{figure}[H]
    \centering
    \begin{subfigure}[b]{\textwidth}
        \centering
        \includegraphics[width=\linewidth]{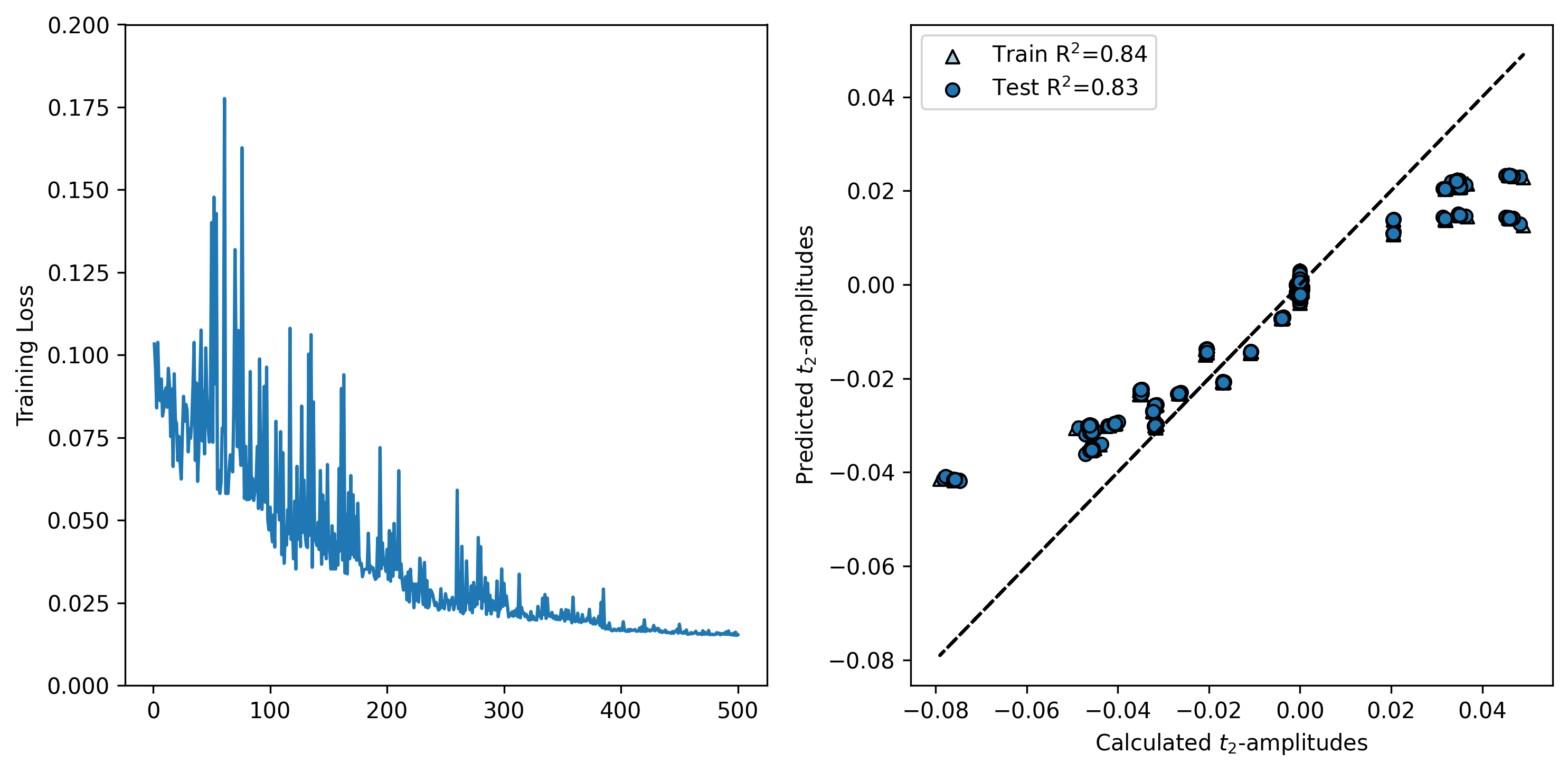}
        \caption{}
        \label{fig:qiskit_statevector_0.1}
    \end{subfigure}
    \hfill
    \begin{subfigure}[b]{\textwidth}
        \centering
        \includegraphics[width=\linewidth]{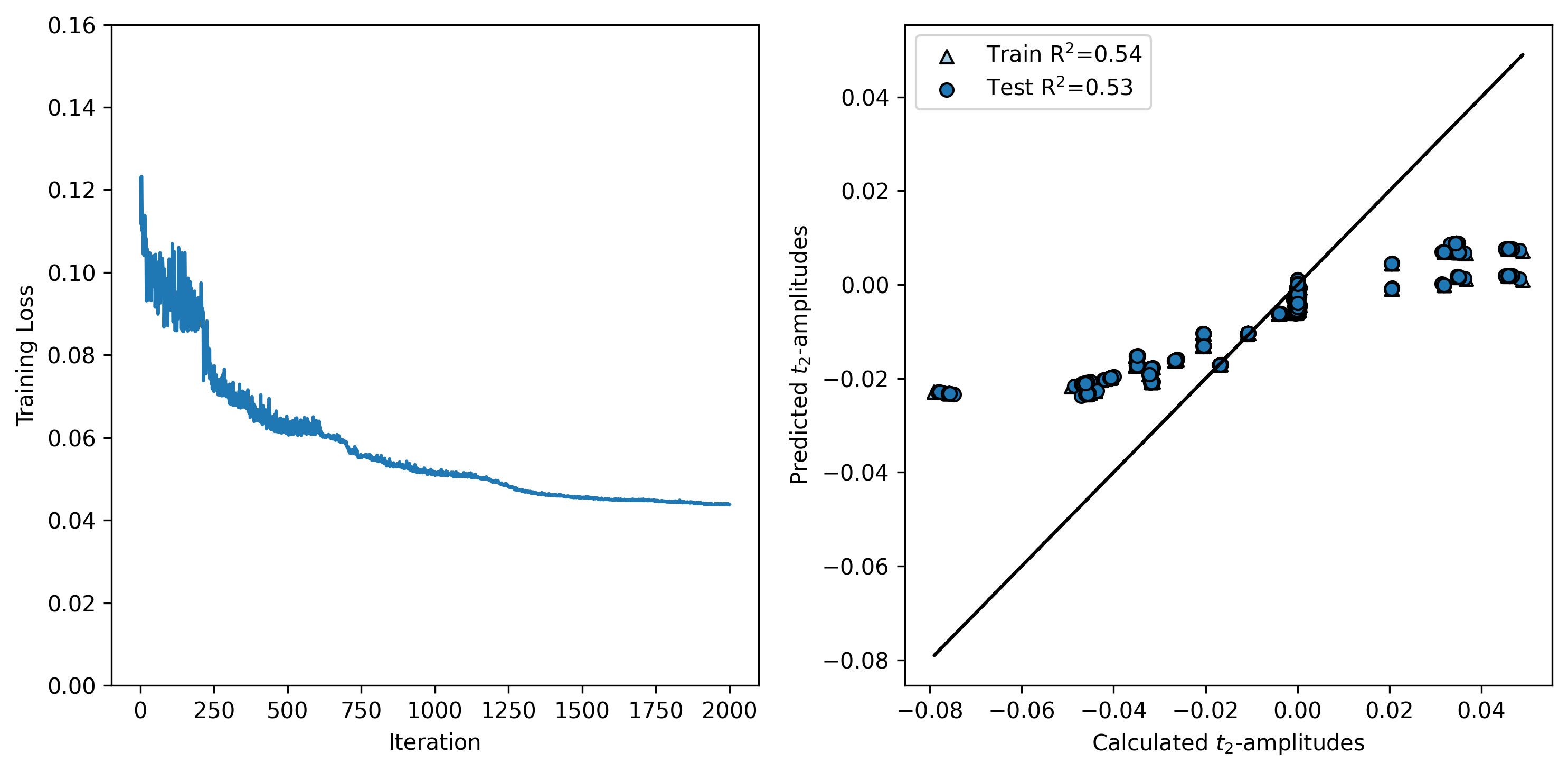}
        \caption{}
        \label{fig:3W_qiskit_statevector_0.1}
    \end{subfigure}
    \caption{State-vector simulation using the Qiskit implementation of A2{\_}HWE-CNOT on the DDCC dataset using (a) the standard approach of one feature per qubit and (b) using two qubits per feature.}
    \label{fig:enter-label}
\end{figure}

\bibliography{achemso-demo}